\documentclass[aps,prb,onecolumn,superscriptaddress,floatfix,showpacs,amsmath,amssymb,nofootinbib,longbibliography,nobibnotes,noeprint]{revtex4-2}

\usepackage{graphicx}
\usepackage{dcolumn,xcolor}
\usepackage{bm}
\usepackage{color}
\usepackage{tabularx}
\usepackage{siunitx}
\usepackage{verbatim}
\usepackage{parskip,stackengine}
\usepackage{soul,xr}
\usepackage[normalem]{ulem}

\newcommand{\rk}[1]{\textcolor{black}{#1}}
\newcommand{\rco}{{\it R}CrO$_3$}
\newcommand{\sco}{SmCrO$_3$}

\newcommand{\rwp}{$R_{\rm wp}$}

\newcommand{\bpa}{$B\parallel {\rm a}$}
\newcommand{\bpb}{$B\parallel {\rm b}$}
\newcommand{\bpc}{$B\parallel {\rm c}$}
\newcommand{\tn}{$T_{\rm N}$}
\newcommand{\tnS}{$T_{\rm {N2}}$}
\newcommand{\tsr}{$T_{\rm {SR}}$}
\newcommand{\tcomp}{$T_{\rm {comp}}$}
\newcommand{\chia}{$\chi_{\rm a}$}
\newcommand{\chic}{$\chi_{\rm c}$}
\newcommand{\chiazfc}{$\chi_{\rm a}^{\rm ZFC}$}
\newcommand{\chiafcw}{$\chi_{\rm a}^{\rm FCW}$}

\newcommand{\mb}{$\mu_{\rm B}$}

\newcommand{\ergGmol}{erg/(G$^2$mol)}

\newcommand{\mbfu}{\(\mu _{\rm B}/{\rm f.u.}\)}

\newcommand{\jmk}{J/(mol\,K)}

\newcommand{\cpp}{$c_{\rm p}$}
\newcommand{\tntwo}{$T_{\rm N2}$}
\newcommand{\tnthree}{$T_{\rm N3}$}

\newcommand{\rv}[1]{\textcolor{black}{#1}}

\begin{document}


\title{The anisotropic magnetic phase diagrams, tricriticality, and spin-reorientation in high-pressure grown SmCrO$_3$ single crystals}

\author{Ning~Yuan}
\affiliation{Kirchhoff Institute of Physics, Heidelberg University, INF 227, D-69120 Heidelberg, Germany}

\author{Erik~Walendy}
\affiliation{Kirchhoff Institute of Physics, Heidelberg University, INF 227, D-69120 Heidelberg, Germany}

\author{Nour~Maraytta}
\affiliation{Institute for Quantum Materials and Technologies, Karlsruhe Institute of Technology, Kaiserstr. 12, 76131 Karlsruhe, Germany}

\author{Waldemar~Hergett}
\affiliation{Kirchhoff Institute of Physics, Heidelberg University, INF 227, D-69120 Heidelberg, Germany}

\author{Luca~Bischof}
\affiliation{Kirchhoff Institute of Physics, Heidelberg University, INF 227, D-69120 Heidelberg, Germany}

\author{Michael~Merz}
\email{michael.merz@kit.edu}
\affiliation{Institute for Quantum Materials and Technologies, Karlsruhe Institute of Technology, Kaiserstr. 12, 76131 Karlsruhe, Germany}
\affiliation{Karlsruhe Nano Micro Facility (KNMFi), Karlsruhe Institute of Technology, Kaiserstr. 12, 76131 Karlsruhe, Germany}

\author{Rüdiger~Klingeler}
\email{klingeler@kip.uni-heidelberg.de}
\affiliation{Kirchhoff Institute of Physics, Heidelberg University, INF 227, D-69120 Heidelberg, Germany}

\date{\today}

\begin{abstract}

SmCrO$_3$ single crystals were successfully grown utilizing the high-pressure optical floating-zone method and their crystal structure, magnetization behavior, and magnetic phase diagrams were thoroughly investigated. Magnetic studies were conducted for fields applied along all principal crystallographic directions, with measurements taken at temperatures as low as 0.4~K and magnetic fields up to 14~T. 
The single crystal growth parameters are reported and the orthorhombic structure with the centrosymmetric space group $Pbnm$ is confirmed. Long-range order of the Cr$^{3+}$ and Sm$^{3+}$ magnetic sublattices evolves at \tn\ = 192~K and \tntwo~=~3~K, respectively. In contrast to previous studies on polycrystals our single crystal data imply a discontinuous and one-step spin-reorientation (SR) of net magnetic moments $\tilde{M}$ from the $c$ axis into the $ab$ plane at zero magnetic field at \tsr~=~33~K. Its discontinuous nature is maintained if $B$ is applied $||c$ axis but tricritical behavior and a triple point is found for $B||a$ axis. While our data are consistent with the magnetic representation $\Gamma_4$ for $T > T_{\mathrm {SR}}$, the size and in-plane direction of the observed net magnetic moment disagree to previously proposed spin configurations, i.e., $\Gamma_1$ and $\Gamma_2$, for the spin-reoriented phases.  
In general, our high-quality single crystals enable us to revisit the phase diagram and to clarify the complex magnetism in \sco\ arising from the interplay of anisotropic 3$d$ and 4$f$ magnetic sublattices. 




\end{abstract}

\maketitle

\section{Introduction}





The interplay between transition metal and rare earth magnetic moments is highly relevant in many research areas including the important application field of permanent magnets such as Sm-Co~\cite{herbst1991r,sagawa1987nd} and Nd-Fe-B systems~\cite{perkins1977interpretation,buschow1977intermetallic,coey2010magnetism}. The phenomena arising especially from the interplay of 4$f$ and 3$d$ magnetic sublattices are particularly complex and give rise to a plethora of intriguing effects 
such as spin reorientation~\cite{kimel2004laser,yamaguchi2013terahertz,Maeter2009,su2010temperature,Popova2007,Popova2008Dy,qian2014study,Stockert2012}, solitonic lattices~\cite{artyukhin2012solitonic}, emerging spin-phonon coupling~\cite{weber2022emerging}, multiferroicity~\cite{tokunaga2012electric,tokunaga2009composite,Adem2010,hassanpour2021interconversion,su2011study,zhou2010intrinsic,sahu2008modification}, spin switching~\cite{yuan2013spin,cao2014temperature} and exchange bias behavior~\cite{huang2014magnetic}. The class of \rco\ studied here crystallizes in a distorted perovskite-type crystal structure with space group (SG) $Pbnm$ (which has the same symmetry as the standardized SG $Pnma$, yet comprises simply an alternative crystallographic setting with permuted principal axes) and two magnetic sublattices ({\it R}$^{3+}$ and Cr$^{3+}$)~\cite{yamaguchi1974theory}. A main feature of the resulting magnetic interactions, 
among 
the 3$d$ moments, among 
the 4$f$ moments, and between 3$d$ and 4$f$ moments, is the evolution of canted antiferromagnetic order of the Cr$^{3+}$ moments, involving a weak ferromagnetic component due to Dzyaloshinskii-Moriya (DM) interactions~\cite{dzyaloshinsky1958thermodynamic,moriya1960anisotropic,moskvin2019dzyaloshinskii}. 

Research on rare earth orthoferrites dates 
back many decades. A review of early works~\cite{hornreich1978magnetic} reports the existence of three spin configurations: $\Gamma_1$ (Cr$^{3+}$: $A_x$, $G_y$, $C_z$; {\it R}$^{3+}$: $O_x$, $O_y$, $C_z$.) spin configuration with no net magnetic moment, while $\Gamma_2$ (Cr$^{3+}$: $F_x$, $C_y$, $G_z$; {\it R}$^{3+}$: $F_x$, $C_y$, $O_z$.) and $\Gamma_4$ (Cr$^{3+}$: $G_x$, $A_y$, $F_z$; {\it R}$^{3+}$: $O_x$, $O_y$, $F_z$.) show net magnetic moments pointing along the $x$ and $z$-axis, respectively (cf. Fig.~S2 of the Supplemental Material (SM)~\cite{SM}). The recent observations of sizable spontaneous polarization in the magnetically ordered phase triggered by the existence of high-pressure grown single crystals has renewed the interest in orthochromite research~\cite{mishra2023structural,rajeswaran2012field,meher2014observation}.\@ Whether {\it R}CrO$_3$ serves 
as a potential multiferroic material remains, however, an ongoing and contentious issue. 
Raman studies on Sm-substituted GdCrO$_3$~\cite{das2023phonon} reveal direct spin-phonon coupling and suggest potential applications for magnetic switching devices. Notably, clear differences appear between single-crystal and polycrystalline samples as demonstrated for ErCrO$_3$~\cite{yano2023magnetic}.\@ For the latter system 
a strong anomaly at the spin-reorientation transition (SRT) and the Er$^{3+}$ magnetic ordering transition in the specific heat is only observed in studies on single crystals. 

Due to the volatilization of Cr\(_2\)O\(_3\) as well as the high melting point of \rco\ compounds, the growth of single crystals is challenging. In this work, \sco\ single crystals were 
successfully grown by the high-pressure optical floating-zone method under Argon (Ar) pressure of 30~bar. The obtained single crystals have been used to investigate the magnetic and thermodynamic properties and to examine 
the effect of external magnetic field applied along the main crystallographic axis. While consistent
with previous studies on polycrystals~\cite{huang2014magnetic,gupta2016study,rajeswaran2012field,sau2021high,sau2022first,tripathi2016phase,tripathi2017evolution,tripathi2019role}, long range magnetic order of the Cr-sublattice evolves at \tn~=~192~K. The successful synthesis of single crystals enables us to study the SRT at \tsr~=~33~K along the principal crystallographic axes $a$, $b$, and $c$ individually. 
Applying the magnetic fields along all three crystallographic axes results in field-induced spin reorientation with critical fields varying with temperature and magnetic field direction and allows to construct the magnetic phase diagrams. We find clear differences with respect to previously reported studies on polycrystals, e.g., by following the uncompensated magnetic moment and its direction in the various magnetically ordered phases. In contrast to previous reports, our single crystal measurements show that the SRT, at $B=0$~T, appears discontinuously and in a one-step transition. In addition, below \tsr , our data do not agree to the previously reported spin configurations. Applying magnetic fields $B||a$ drives the system towards a triple point and tricritical behavior. Distinct anomalies in the magnetization and specific heat signal the appearance of long-range magnetic order of Sm$^{3+}$ moments at \tnS~=~3~K. 

\section{Experimental methods}

Polycrystalline \sco\ was synthesized by a standard solid-state reaction. Stoichiometric amounts of Sm\(_2\)O\(_3\) (99.9~\%, Alfa Aesar) and Cr\(_2\)O\(_3\) (99.6~\%, Alfa Aesar), along with a 10~\% excess of Cr\(_2\)O\(_3\), were thoroughly mixed in a mortar and calcined at 1350~$^\circ$C for 48~h (air flow, ambient) with several intermediate grindings~\cite{MASSA2018294,schneider1961solid}. The obtained powder was reground, packed in a rubber tube and isotropically pressed at 60~MPa to produce cylindrical rods with a length of 5-6~cm and a diameter of 5~mm. The rods were then annealed for 48~h (air flow, ambient) at 1500~$^\circ$C. Single crystals of \sco\ were successfully grown using the high-pressure optical floating-zone furnace (HKZ, SciDre)~\cite{neef2017high}. A 7~kW Xenon arc lamp was used as the heat source, and a 30~bar Ar atmosphere was maintained with an Ar flow of 0.1~l/min. The feed and seed rods were pulled at a rate of 10~mm/h. To improve the homogeneity of the melting zone, counter-rotation of the feed and seed rods at 10~rpm was necessary. Further details on the growth and the sample characterization can be found in Ref.~\cite{NingDiss}.

The phase purity and crystallinity of the resulting materials were studied by powder X-ray diffraction (PXRD) and the Laue method in back-scattering geometry. 
PXRD was performed at room temperature by means of a Bruker D8 Advance ECO diffractometer using Cu-K$\alpha$ radiation ($\lambda$ = 1.5418~\AA). The data have been collected in the 2$\Theta$ range of 10 – 90$^\circ$ with 0.02$^\circ$ step-size. Laue diffraction was done on a high-resolution X-ray Laue camera (Photonic Science). 

The spatial structure of the prepared sample was studied by single crystal x-ray diffraction (SC-XRD) using Mo K$_{\alpha}$ radiation and a resolution of 0.5 \AA. Temperature-dependent SC-XRD measurements were performed at 300 and 80~K on a high-flux, high-resolution rotating anode RIGAKU Synergy-DW (Mo/Ag) system. The diffractometer is equipped with pairs of precisely manufactured Montel mirror optics, a motorized divergence slit which was set to 5 mrad for these measurements, and a background-less Hypix-Arc150$^{\circ}$ detector which guarantees the lowest reflection profile distortion and ensures that all reflections are detected under equivalent conditions. For all temperature steps nearly 12000 Bragg peaks were collected, significantly reducing the experimental uncertainty. The investigated sample shows no mosaic spread or additional reflections from secondary phases which demonstrates the high quality of the sample and allows for an excellent evaluation with the most recent version of the CrysAlisPro software package~\cite{CrysAlis}. The data were corrected for Lorentz, polarization, extinction, and absorption effects. Using SHELXL~\cite{Sheldrick1,Sheldrick2} and JANA2006~\cite{Vaclav_229_2014}, all averaged symmetry-independent reflections (I $>$ $2\sigma$) have been included for the refinements. For both measured temperatures, the unit cell and the SG were determined, the atoms were localized in the unit cell using random phases 
methods, the structure was completed and solved using difference Fourier analysis, and finally the structure was refined.

The magnetization has been measured in a SQUID (superconducting quantum interference device) magnetometer (MPMS3, Quantum Design Inc.) and in a Physical Properties Measurement System (PPMS-14, Quantum Design Inc.) employing the vibrating sample magnetometer option. A relaxation method was used to perform specific heat measurements in the PPMS. When studying the temperature dependence of magnetization, zero field cooled-warming (ZFC), field cooled-cooling (FCC) and field cooled-warming (FCW) protocols were applied where the sample has been cooled down to the lowest temperature either at zero field (ZFC) before applying the external magnetic field, or in the actual measurement field (FC). FC measurements have been done upon warming (FCW) or cooling (FCC). Before the ZFC measurements, the applied magnetic field was zeroed-out by using the oscillation mode at room temperature.

\section{Results}
\subsection{Single crystal growth and structure refinement}

While multiple studies have been reported on the preparation of polycrystalline \sco, to the best of our knowledge, no oriented \sco\ single crystals have been reported which may be attributed to the challenges of heavy volatilization of Cr\(_2\)O\(_3\) and the high melting point of \sco. In many systems, the floating-zone (FZ) method is considered to have an advantage in increasing the size and quality of single crystals compared to flux methods while volatilisation is known to be suppressed by high Ar-pressure~\cite{Wizent2011,Boothroyd2011,hergett2019high,koohpayeh2008optical,Schmehr2019}. This led us to investigate 
the crystal growth conditions for \rco\ under high pressure. While the growth under high pressure can effectively mitigate the volatilisation of Cr\(_2\)O\(_3\), the required temperature at which the material melts will increase with pressure, 
approaching the operational limitation of the device. During the successful growths, in-situ temperature measurements by means of a two-color pyrometer~\cite{dey2019magnetic,hergett2019high} determined the temperature of the melting zone to about 2300~$^\circ$C.

\begin{figure}[tb]
\includegraphics[width=\columnwidth,clip]{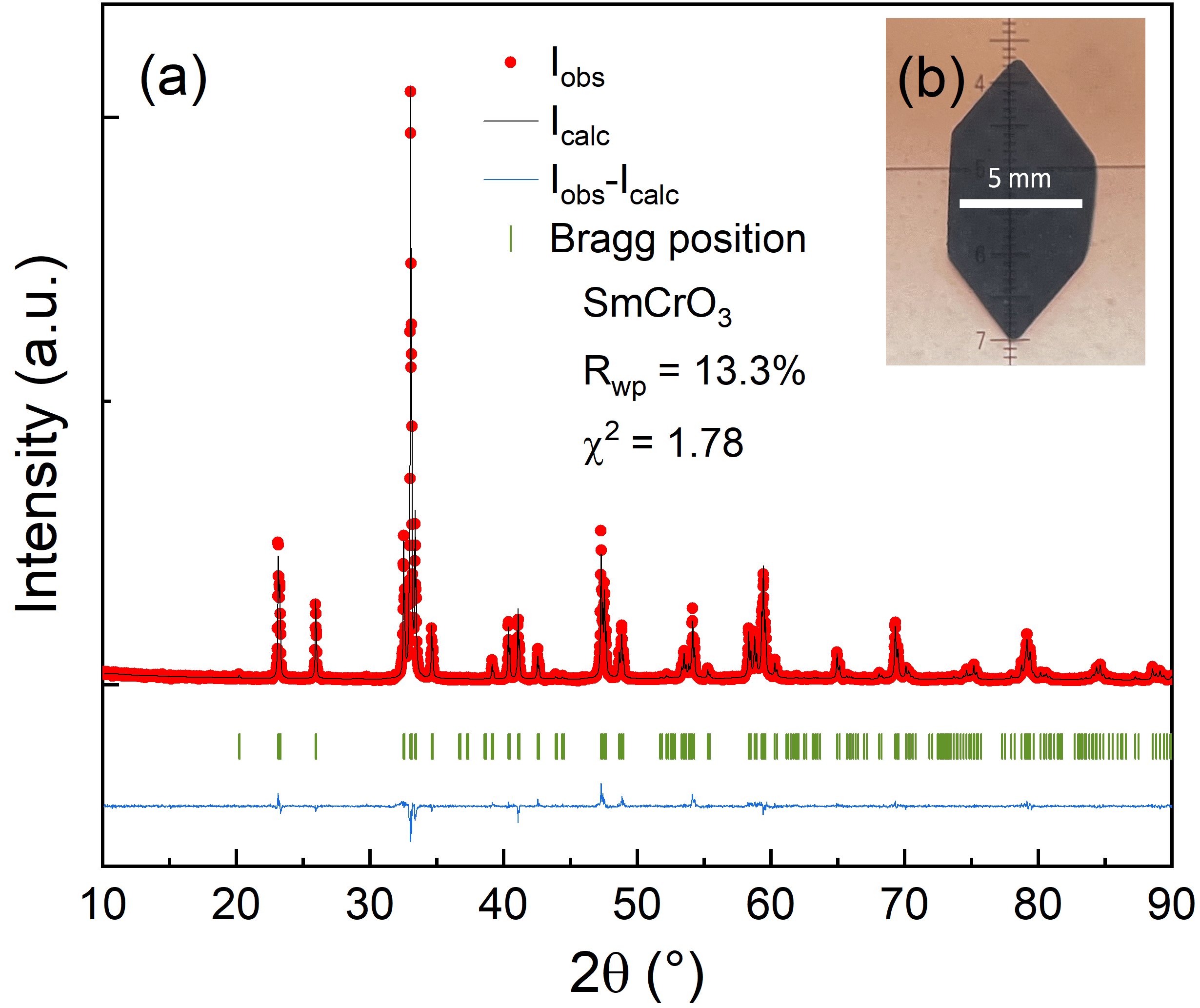}
\caption{(a) Room temperature powder XRD patterns and corresponding Rietveld refinement~\cite{rodriguez2001introduction} of ground \sco\ single crystals. The observed diffraction pattern is shown in red, the calculated one in black, and the difference between them is shown in blue. The vertical green bars show the expected Bragg positions of \sco\ (ICSD No.~5988~\cite{sco-xrd2017evolution}). The refinement converged to \rwp\ = 13.3 $\%$ and $\chi^{2}$ = 1.78. (b) Picture of oriented single crystal used for thermodynamic measurements.}
\label{SCO_XRD}
\end{figure}

\begin{figure}[ht]
\centering
\includegraphics[width=\columnwidth,clip]{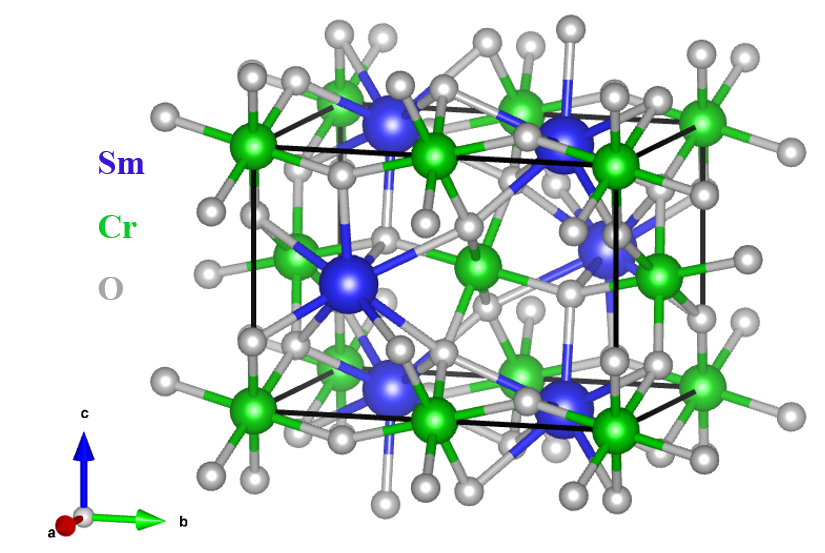}
\caption{Orthorhombic spatial structure of SmCrO$_3$ at 300 K. Thin black lines represent the orthorhombic unit cell.}
\label{fig:crystal structure of SmCrO3}
\end{figure}

We also observed significant amounts of deposited  Cr\(_2\)O\(_3\) volatiles adhering to the inner protection glass tube, thereby affecting the focusing of light (see Fig.~S1a of the SM~\cite{SM}). Eventually it was determined by our experiments that the corresponding single crystals could be prepared at 30~bar Ar atmosphere. A relatively fast growth rate of 10~mm/h was chosen in order to further reduce 
Cr\(_2\)O\(_3\) volatilization. X-ray Laue diffraction in back scattering geometry was used to confirm single crystallinity and to orient the single crystals which were then cut with respect to the crystallographic {\it a}, {\it b} and {\it c}-axis using a diamond-wire saw (see Fig.~S1b and c of the SM~\cite{SM}). The powder x-ray diffraction (PXRD) on the ground \sco\ single crystals as well as the Rietveld refinement to the powder data are shown in Fig.~\ref{SCO_XRD}. The result of the PXRD refinement demonstrates that the sample is free of impurity phases, and the lattice parameters and the crystal structure are consistent with previous reports~\cite{su2012dependence,tripathi2019role,sau2021high}. Refined structural parameters are shown as Table~I in the SM~\cite{SM}.



The quality and structure of our single crystals have been further determined by SC-XRD measurements. For the crystal structure, two orthorhombic space groups (SGs) were suggested in the literature and can be confirmed from our refinements: The centrosymmetric SG $Pbnm$ (or the standardize SG $Pnma$ see discussion above and \cite{sau2021high} as well as \cite{sco-xrd2017evolution}), and the non-centrosymmetric SG $Pbn2_1$ (which is the crystallographic subgroup of $Pbnm$ and simply an alternative setting of the standardized SG $Pna$2$_{1}$) (\cite{ding_102_2029} and \cite{ghosh_107_2014}). For both measured temperatures the structure was refined in these two relevant SGs.\@ The corresponding results of the refinements are illustrated in Table~II in the SM~\cite{SM}. The displacement parameters were refined anisotropically, however, due to space limitations only the equivalent ones $U_{\rm equiv}$ are listed in the Table~II. Errors shown are statistical errors from the refinement. For the 300~K as well as for the 80~K data, the structural refinements converged quite well for both suggested SGs and showed excellent reliability factors (see Table~II in SM~\cite{SM} for $wR_\mathrm{2}$, $R_\mathrm{1}$ and GOF values). As a consequence, both proposed SGs are in principle possible. Anyhow, the higher symmetry SG $Pbnm$ shows slightly better agreement factors despite a smaller number of parameters for the atomic positions compared to the non-centrosymmetric one. Furthermore, when reducing the symmetry from $Pbnm$ to $Pbn$2$_1$ the inversion symmetry is lost, however, has then to reappear in the form of a twinning element which produces two merohedral inversion twins. Therefore, the so-called Flack parameter, $f$, is introduced to estimate the absolute configuration of a non-centrosymmetric structural model and, thus, the respective percentage of the two inversion twins has to be refined which further increases the number of used parameters for the latter SG. The refinements of the absolute structure for SG $Pbn$2$_1$, indicate that the fraction is around $f = 50$~\% for both twins which points to the fact that the sample is either perfectly twinned or that the SG remains centrosymmetric. By taking all of the above mentioned factors into account and unless no other experiments 
unequivocally prove that the compound is acentric, it is generally accepted convention to use the higher symmetry and, hence, SG $Pbnm$ has to be favored. The non-centrosymmetric refinement in SG $Pbn$2$_{1}$ then only comes at the cost of introducing more degrees of freedom (i.~e., more parameters) and even results in slightly worse agreement factors.
\rk{Note, that the analysis of the XRD patterns also implies that the ratio Sm:Cr:O3 (with the Sm occupation fixed to 1) very precisely amounts to 1.00:1.00:3.00, with an error in the per mille range for Sm and Cr, and less than 1 percent for the oxygen.}

\subsection{Magnetic order and temperature-driven spin-reorientation \label{ch:srt}}

\begin{figure}[htb]
\includegraphics[width=\columnwidth,clip]{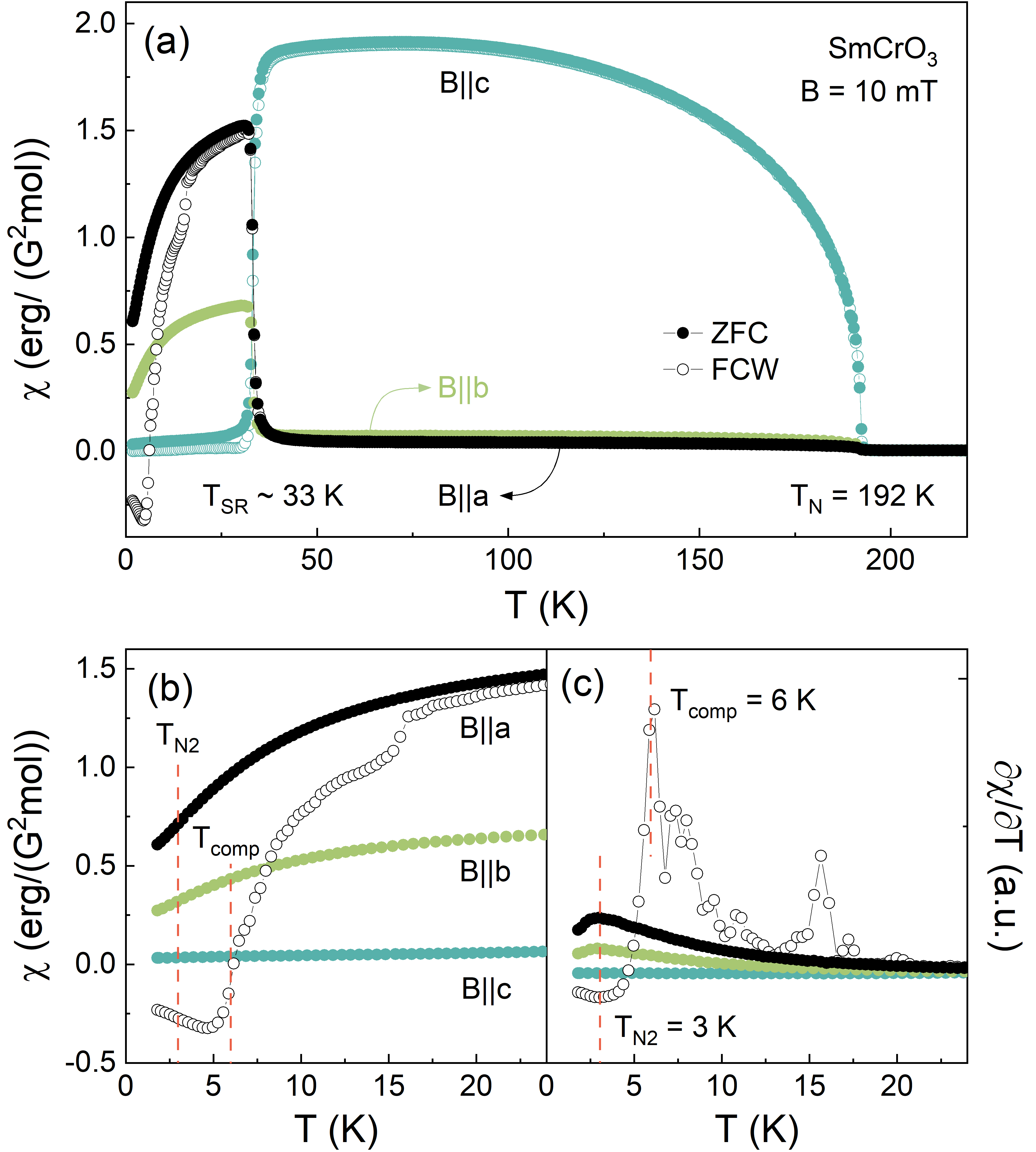}
\caption{(a) Temperature dependence of the static magnetic susceptibility $\chi = M/B$, obtained at $B=10$~mT applied along the crystallographic $c$ axis (\bpc), $b$ axis (\bpb), and $a$ axis (\bpa); ZFC and FCW data are represented by open and solid circles, respectively. (b) Static magnetic susceptibility and (c) its derivative for $B||c$, $B||b$, and $B||a$ at $B=10$~mT and at $T \leq 30$~K. \tn, \tnS, and \tcomp\ have been determined as described in the text.}\label{SCO_MT}
\end{figure}

The \sco\ single crystal reveals a pronounced magnetic anisotropy as displayed in the temperature dependence of the static magnetic susceptibility $\chi(T)=M(T)/B$, obtained at $B=10$~mT, which is shown in Fig.~\ref{SCO_MT} for fields applied along the different crystallographic axes. The onset of long-range antiferromagnetic order at \tn~=~192~K (the phase will be labelled AF I) is associated with a pronounced increase in \chic\ signalling the presence of a sizable net moment along the $c$ axis as expected for canted antiferromagnetic order and in-line with previous studies on polycrystals~\cite{tripathi2016phase,qian2014study,rajeswaran2012field}. At high temperatures, the variation of magnetic susceptibility with temperature presents a Curie-Weiss-like behavior as demonstrated by $\chi_{\rm c}^{-1}(T)$ (red line) in Fig.~S3 of the SM~\cite{SM}. 
Fitting the data obtained at 1~T above 370~K by a Curie-Weiss law yields an effective moment $p_{\rm eff} = 3.5(1)$~$\mu_{\rm B}$ and a Weiss temperature $\Theta_{\rm{W}} = -373(3)$~K.~\footnote{\rv{The measurement field was chosen to optimize the signal-to-noise ratio while observing the low-field limit of $B\ll B_{\rm sat}~\sim 200$~T estimated below.}} The obtained $p_{\rm eff}$ is lower than those reported in previous studies on polycrystalline samples~\cite{qian2014study,sau2021high,sau2022first},
but closer to the theoretical value of 3.96~$\mu_{\rm B}$ given by $p_{\rm eff} = \sqrt{p_{\rm{Sm}}^{2}+p_{\rm{Cr}}^{2}}$~\footnote{The theoretical value has been estimated by assuming $p_{\rm eff}= 0.845$~$\mu_{\rm B}$ for the free-ion paramagnetic contribution of the Sm$^{3+}$ moments ($J = 5/2$) and $p_{\rm eff}= 3.873$~$\mu_{\rm B}$ for Cr$^{3+}$ moments ($J = 3/2)$.}. The slightly reduced value of $p_{\rm eff}$ might suggest that crystal fields are still relevant at 400~K, thereby affecting the Sm$^{3+}$ moments which $^{6}H_{5/2}$ ground state is likely split in three doublets (see, e.g., the discussion in Ref.~\cite{MASSA2018294}). In addition, the value of $\Theta_{\rm{W}}$ from fitting our data $\chi(T,B=1~{\rm T})$ above 370~K might indicate limited validity of our analysis in terms of the Curie-Weiss model which is only correct in the high temperature regime $T>\Theta_{\rm W}$. Our obtained $\Theta_{\rm W}$ is in particular much smaller than those reported in the literature obtained at smaller fields and lower temperatures, i.e., $\Theta_{\rm W}=-880$~K (from a fit to $\chi(B=0.1~{\rm T},~T<300~{\rm K})$~\cite{qian2014study}; $-1326$~K (from $\chi(0.05~{\rm T},~210-250~{\rm K})$~\cite{sau2021high}; -1002~K (from $\chi(0.15~{\rm T},~210-250~{\rm K})$~\cite{sau2022first} \rv{which may be explained by CF effects in the reported lower temperature regimes.} For comparison, we have also fitted our data by means of a Moriya model (see Refs.~\cite{moriya1960anisotropic,mcdannald2015magnetic,yin2017magnetic} and Fig.~S3 of the SM~\cite{SM}) which yields $p_{\rm eff} = 3.6(1)$~$\mu_{\rm B}$, $\Theta_{\rm{W}} = -435(2)$~K, $J/k_{\rm B} = 12.84(1)$~K, and $D/k_{\rm B} = 1.54(2)$~K, the latter being the symmetric and antisymmetric exchange interactions of the Cr$^{3+}$ moments. 

Upon cooling below \tn , \chic\ reaches a maximum value of 1.91(1)~\ergGmol\ at $\simeq 70$~K and then decreases sharply at
\tsr~$\simeq 33$~K. In contrast, \chia\ and $\chi_{\rm b}$ rapidly increase at \tsr . The data suggest that the net magnetic moment aligns along the $c$ axis for \tsr $< T <$ \tn\ but rotates from the $c$ axis into the $ab$ plane for $T <$ \tsr. Very small anomalies in $\chi_{\rm a}$ and $\chi_{\rm b}$, at \tn\ (cf. Fig.~\ref{SCO_MT} and the SM~\cite{SM}), are in the range of uncertainty of the crystal orientation and the cutting process so that our data, for \tsr~$< T <$~\tn, are in agreement with the net spin component fully aligned along the 
$c$ axis, i.e., with the $\Gamma_4$ configuration. 

Figure~\ref{SCO_MT}b illustrates a pronounced hysteresis between \chia\ obtained in the FC and the ZFC protocols which further confirms the presence of a 
weak ferromagnetic component in the $ab$ plane at $T\leq$~\tsr . 
Particularly, \chiazfc\ becomes zero at a compensation point around \tcomp~=~6~K and has negative values at lower temperature. We attribute this to the interplay of the Sm$^{3+}$ and Cr$^{3+}$ magnetic sublattices. At \tcomp, the magnetic moments of both sublattices are of equal magnitude but opposite directions, thereby canceling each other out 
and leading to zero magnetization and static magnetic susceptibility $\chi$. The persistent decrease of \chiafcw\  in the low-temperature region suggests the ordering of Sm$^{3+}$. This is corroborated by a peak in $\partial \chi_{\rm a}/\partial T$ at \tnS~=~3~K (see Fig.~\ref{SCO_MT}c and the discussion in §~\ref{chap:tntwo}). The low-temperature tail of \chiazfc\ also suggests the antiparallel arrangement of Sm$^{3+}$ and Cr$^{3+}$ magnetic moments. In Fig.~S5 of the SM~\cite{SM}, the temperature hysteresis in \chia\ and \chic\ is shown for various higher magnetic fields up to 0.2~T which demonstrates the interplay of external fields on bifurcation. 

\begin{figure}[hbt]
\includegraphics[width=0.9\columnwidth,clip]{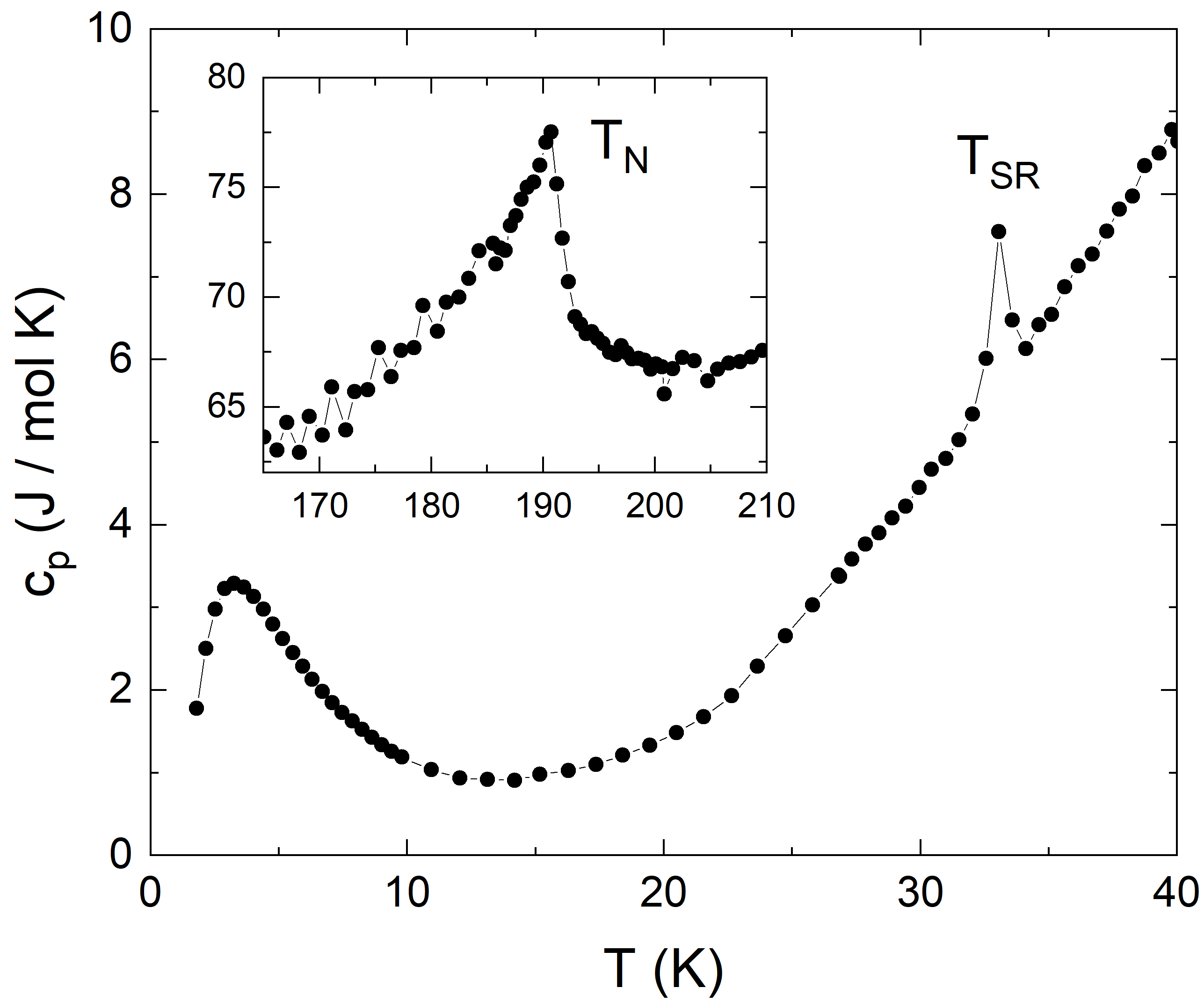}
\caption{Specific heat of SmCrO$_3$ at $B=0$~T. \tn\ and \tsr\ label the onset of long-range AF order (i.e., phase AF I) of Cr$^{3+}$ moments and SR into the phase AF II.}
\label{SCO_Cp0T}
\end{figure}

The specific heat of the \sco\ single crystal shown in the Fig.~\ref{SCO_Cp0T} verifies the evolution of a long-range magnetically ordered phase (AF I) at \tn\ and confirms a phase transition at \tsr\ into AF II. The observed anomaly at \tn~=~192~K is $\lambda$-shaped which corroborates the continuous nature of the phase transition. A much sharper peak-like anomaly appears at \tsr~=~33~K and its predominantly symmetric shape indicates the first-order nature of the SRT. As will be shown below, the magnetic field $B||c$ effect on \tsr\ can be quantitatively described by the Clausius-Clapeyron equation which is valid only for discontinuous transitions and hence confirms our assignment of the nature of the transition. 
Finally, a broad peak in the specific heat centered around 3~K suggests a Schottky contribution to \cpp. 

A quantitative estimate of the anomaly at \tsr\ implies that SRT is associated with a small jump in entropy $\Delta S_{\rm {SR}} = 0.047(1)$~\jmk . At \tn, one may use the anomaly size of 5.6(25)~\jmk\ as an upper limit of the actual (mean-field) specific heat jump $\Delta c_{\rm p}$ since it may be superimposed by critical fluctuations. However, even this value is much smaller than the expected mean-field value for a $S=3/2$ system of $\Delta c^{\rm mf}_{\rm p} = R\frac{5S(S+1)}{S^2+(S+1)^2} \simeq 18.3$~\jmk , with $R$ being the gas constant~\cite{Barron}. This discrepancy implies significant short-range magnetic order above \tn\ and/or incomplete long-range order below \tn. 

\begin{figure}[hbt]
\includegraphics[width=0.9\columnwidth,clip]{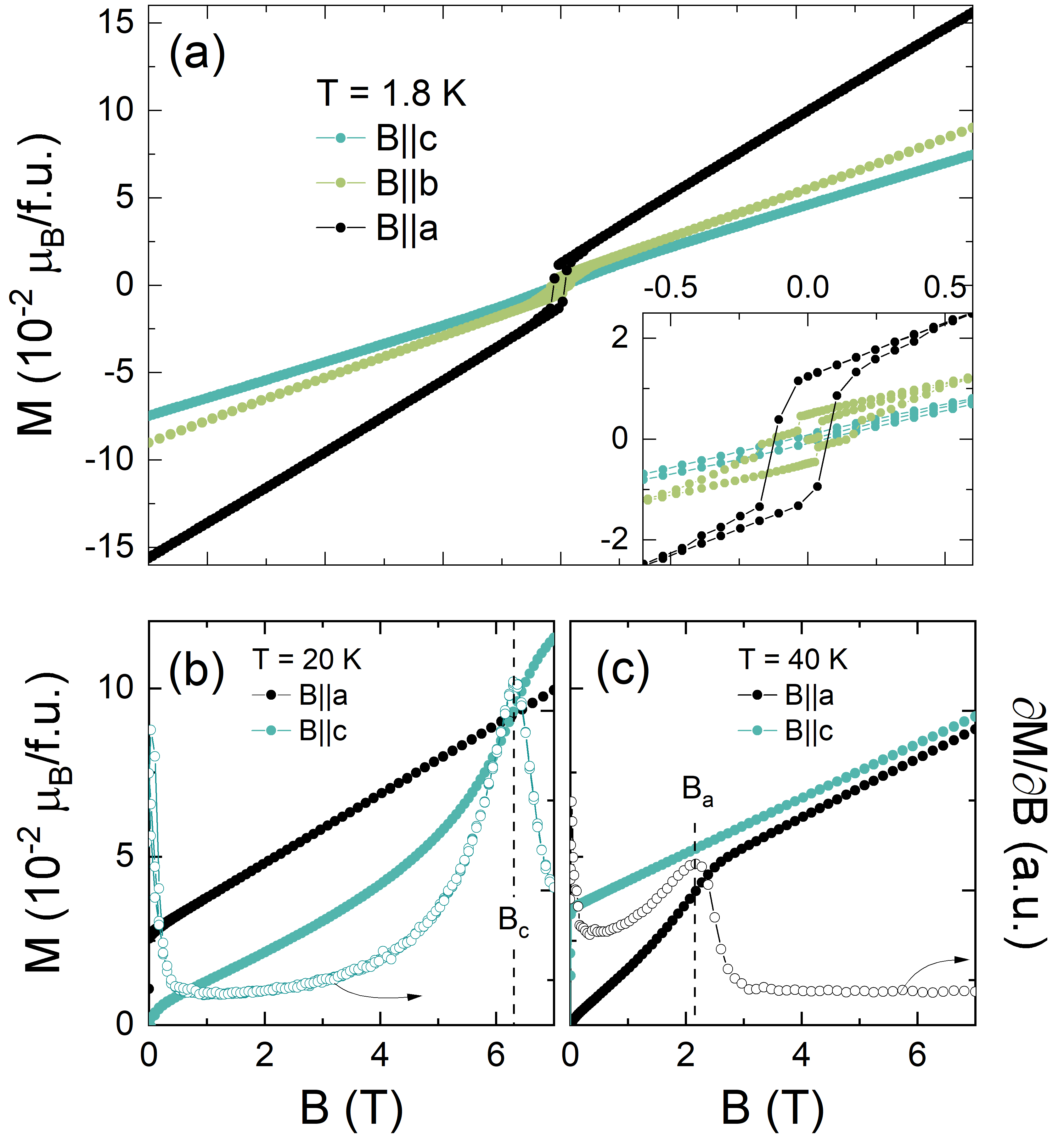}
\caption{(a) Isothermal magnetization at $T = 1.8$~K for $B||a$, $b$, and $c$ axis. The inset highlights the behavior around zero field. (b) Isothermal magnetization at $T = 20$~K for $B||c$ (left ordinate) and corresponding magnetic susceptibility $\partial M/\partial B_{||c}$ (right ordinate), and (c) the same quantities at $T = 40$~K for $B||a$. The dashed lines indicate the critical field $B_{\rm c}$ and $B_{\rm a}$ as described in the text.}\label{SCO_MB}
\end{figure}

The rotation of a small ferromagnetic moment from the $c$ axis towards the $ab$ plane is confirmed by isothermal magnetization data in Fig.~\ref{SCO_MB}. At $T=1.8$~K, $M_{\rm a}$ ($M_{\rm b}$) evidences a small remanent moment of 0.013~\mbfu (0.005~\mbfu) and a hysteresis of about 0.1~T (0.14~T). In contrast, no sizable net moment is seen in $M(B||c)$, which confirms the rotation of the net magnetic moment perpendicular to the $c$ axis at \tsr . The response of the net moment is superimposed by a linear-in-field increase of $M$ which implies AF behavior. At $B=7$~T field, $M_{\rm c}<M_{\rm b}\simeq M_{\rm a}/2$, all being much smaller than the saturation magnetization. Extrapolating $M_{\rm a}$ suggests a saturation field $B||a$ of nearly 200~T.

At $T=20$~K, there is a step-like feature in $M(B||c)$ which is signalled by a peak in $\partial M/\partial B_{||c}$ at $B_{\rm c}=6.4(1)$~T (see Fig.~\ref{SCO_MB}b). While such a feature is not present for the other field directions at 20~K, it appears at $T=40$~K in $M(B||a)$, at $B_{\rm a}=2.3(1)$~T, but at this temperature it is quenched in $M(B||c)$ (Fig.~\ref{SCO_MB}c). We attribute the step-like increase in magnetization to field-induced SRT from the $ab$ plane to the $c$ axis in the SRT phase (AF II) at $T<T_{\rm SRT}$. Conversely, the field-driven rotation appears from the $c$ to the $a$ direction in the high-temperature phase AF I at $T>T_{\rm SRT}$. 

\begin{figure}[tb]
\includegraphics[width=0.9\columnwidth,clip]{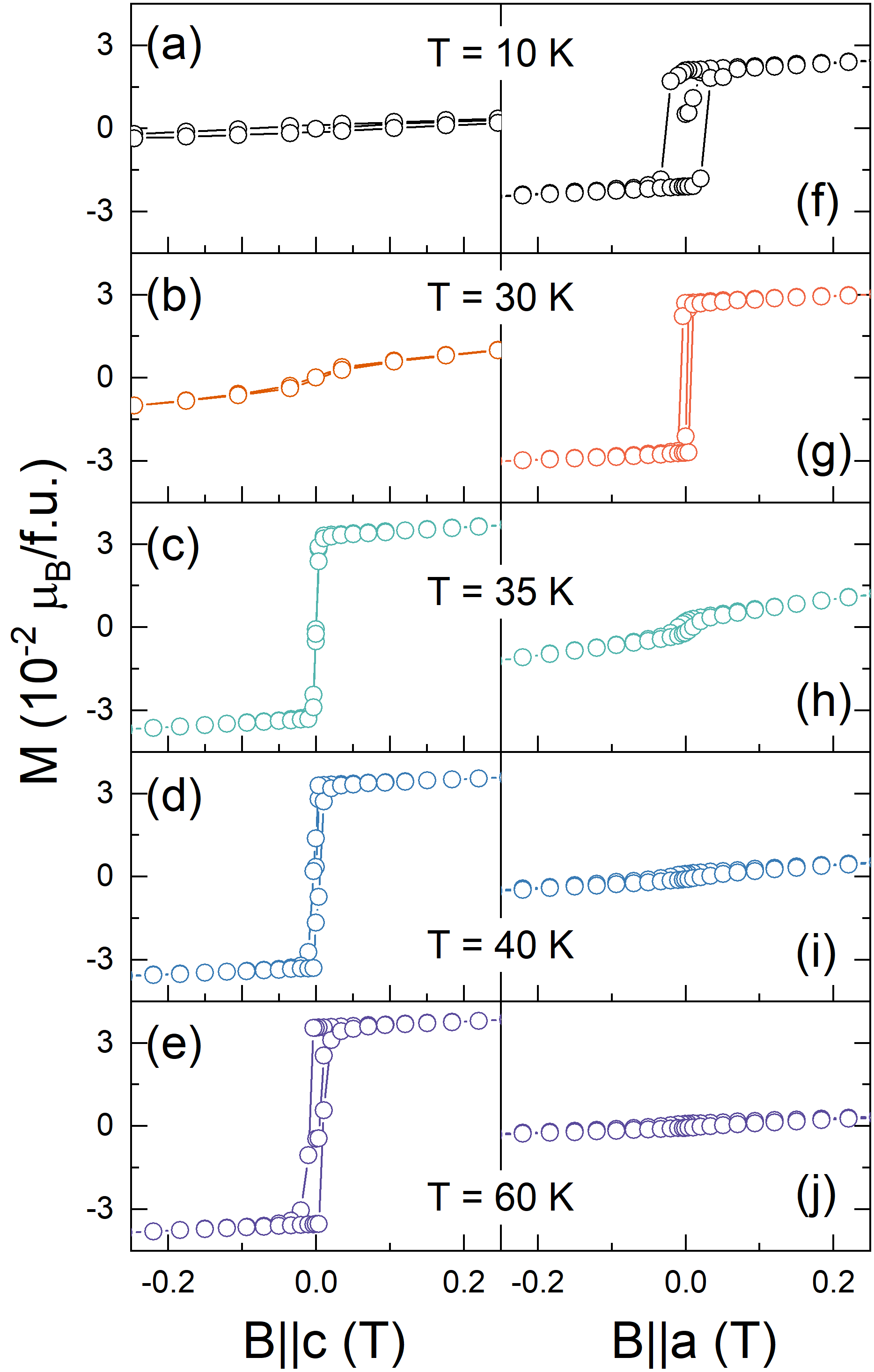}
\caption{Isothermal magnetization at selected temperatures for (a-e) $B||c$ axis and (f-j) $B||a$ axis. Data for $B||b$ are shown in the SM~\cite{SM}.}
\label{SCO_MBall}
\end{figure} 

\begin{figure}[hbt]
\includegraphics[width=0.9\columnwidth,clip]{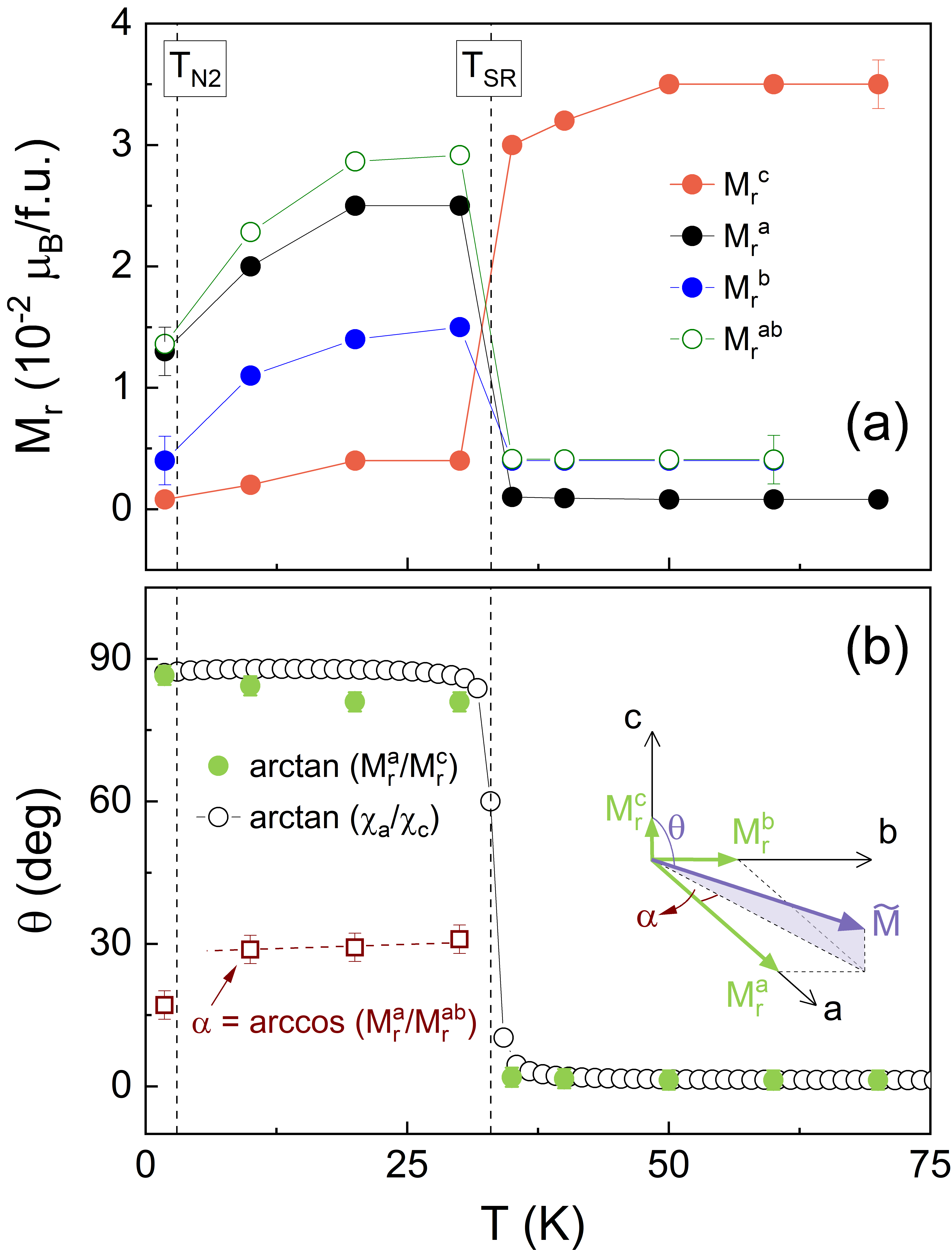}
\caption{(a) Temperature dependence of the remanent magnetic moment $M_{\rm r}^{\rm i}$ along the three crystallographic axes $i=a,b,c$ read-off from Figs.~\ref{SCO_MBall} and S7 of the SM, and calculated in-plane net moment $M_{\rm r}^{\rm ab}=\sqrt{(M_{\rm r}^{\rm a})^2+(M_{\rm r}^{\rm b})^2}$). Error bars include systematic uncertainties associated with the crystal orientation.
(b) Rotation angle $\theta$ calculated from $M_{\rm r}^{\rm a}$ and $M_{\rm r}^{\rm c}$ (green circles) and from the static magnetic susceptibility $\chi = M/B$ obtained at $B = 0.01$~T for $B||a$ and $B||c$ (black circles). The sketch illustrates $\alpha$ (shown by open squares), $\theta$ and $\tilde M$. Vertical dashed lines mark the SRT and \tntwo .
}
\label{SCO_Theta}
\end{figure}

In order to further elucidate the reorientation process, we have performed a series of $M$ vs. $B$ measurements for $B||c$ and for $B||a$, respectively, at different temperatures (see Fig.~\ref{SCO_MBall}). 
Similar to the trend observed at 1.8~K in Fig.~\ref{SCO_MB}, the data for $B||c$, at 10~K display rather linear behavior typically expected for antiferromagnets at low magnetic fields. While a rather rectangular open hysteresis loop is observed for $T >$ \tsr , no such hysteresis is observed  at $T \lesssim$~ \tsr. There is however a tiny hysteresis-free s-shape of the curve $M(B||c)$ in the field range $|B|\leq 0.2$~T at $T \lesssim$~ \tsr, which points to a small magnetic moment and suggests a derivation of the pure $\Gamma_2$ configuration. 
Concomitantly, a weak ferromagnetic open hysteresis curve appears for $B||a$. From the data in Figs.~\ref{SCO_MBall} and S7 of the SM~\cite{SM} we extract remanent magnetic moments $M_{\rm{r}}^{\rm{a}}$, $M_{\rm{r}}^{\rm{b}}$, and $M_{\rm{r}}^{\rm{c}}$ which clearly illustrate the rotation of the weak net moments at the SRT (see Fig.~\ref{SCO_Theta}a). In particular, our single crystal data imply the rotation of the net moment into the $ab$ plane. 

While, for $T >$~\tsr\ (AF I) , the measured $\chi_{\rm c}\times B$ is approximately equal to the total net magnetic moment $\tilde M$ in the system, for $T <$~\tsr\ (AF II) the measured values $\chi_{\rm a}\times B$ ($\chi_{\rm b}\times B$) can be considered as the projection of $\tilde M$ on the $a$ ($b$) axis. The fact that we observe finite $M_{\rm r}^{\rm a}$ and $M_{\rm r}^{\rm b}$ implies that $\tilde M$ is not aligned along one of the crystallographic axes. From our data the angle $\alpha$ between the $a$ axis and $\tilde M$ just below \tsr\ can be estimated to be approximately $30^{\circ}$. It visibly decreases only below \tntwo . Furthermore, the temperature dependencies of $M_{\rm r}^{\rm a}$ and $M_{\rm r}^{\rm c}$ allow us to calculate the temperature dependence of the rotation angle $\theta$, thereby further illustrating the rotation of the net magnetic moments by utilizing the formula~\cite{bazaliy2004spin}:

\begin{equation}
    \theta =\arctan \left ( \frac{M_{\rm{r}}^{\rm{a}}} {M_{\rm{r}}^{\rm{c}}} \right ).
    \label{Eq_SCO_theta}
\end{equation} 

The resulting temperature dependence is shown in Fig.~\ref{SCO_Theta}b. In addition, $\theta$ can be also estimated from $\chi_{\rm c}$/$\chi_{\rm a}$ as shown in Fig.~\ref{SCO_Theta}b, too. The two estimates agree perfectly to each other if the bare remanent moment is considered (data not shown). The slight deviations in Fig.~\ref{SCO_Theta}b (open/filled circles around 25~K) are directly associated to the fact that the small magnetic moments indicated by the s-shaped behavior in $M$ vs. $B$ are naturally not captured by the low-field static magnetic susceptibilities $\chi_a$ and $\chi_c$. As sketched in the inset of Fig.~\ref{SCO_Theta}b, the angle $\theta$ characterizes the rotation of the magnetic moment in the $ac$ plane, whereas the actual rotation plane of $\tilde M$ might be approximated by the purple plane illustrated in Fig.~\ref{SCO_Theta}b. Thus, $\theta= 90^{\circ}$ merely represents the alignment of the magnetic moment in the $ab$ plane. Taken together, the weak ferromagnetic moments are predominantly aligned along the $c$ axis when $T >$~\tsr\, implying the spin configuration $\Gamma_4$. For $T <$~\tsr\, the magnetic moments rotate from the $c$ axis to the $ab$ plane which is consistent with the $\Gamma_2$ phase. The completely 
i.e., in-plane, net moment agrees very well to the net moment $\|c$ just above \tsr\ (Fig.~\ref{SCO_Theta}a). 
The presence of such an uncompensated moment is likely attributed to a canting of the predominantly 
antiferromagnetic spin arrangement and in particular rules out the collinear antiferromagnetic structure $\Gamma_1$. On the other hand, the data also suggest a tiny remaining moment $\|c$ as demonstrated by the s-shaped behavior of $M(B||c)$, thereby indicating deviations of $\Gamma_2$. Moreover, the in-plane net moment direction deviates by $\sim 60^\circ$ from the $b$ axis. These observations do not agree to the pure $\Gamma_2$ case. Hence, our data imply the necessity to reassess the recent controversy whether, at low temperatures, \sco\ exhibits the collinear spin configuration $\Gamma_1$ (Ref.~\cite{tripathi2017evolution}) or $\Gamma_2$ (Ref.~\cite{sau2021high,Mishra2024}). While our data clearly rule out $\Gamma_1$ even at low temperatures, the results are also inconsistent with the pure $\Gamma_2$ representation. Additional measurements, such as neutron diffraction on single crystals, are required to clarify this issue. In addition, our single crystal data imply that the in-plane direction of $\tilde{M}$ is in between $a$ and $b$ but rotates slightly (by $\sim 10^\circ$) when long-range order of the Sm$^{3+}$ finally evolves (see Fig.~\ref{SCO_Theta}b).

\subsection{Magnetic phase diagrams}

\begin{figure}[tb]
\includegraphics[width=0.9\columnwidth,clip]{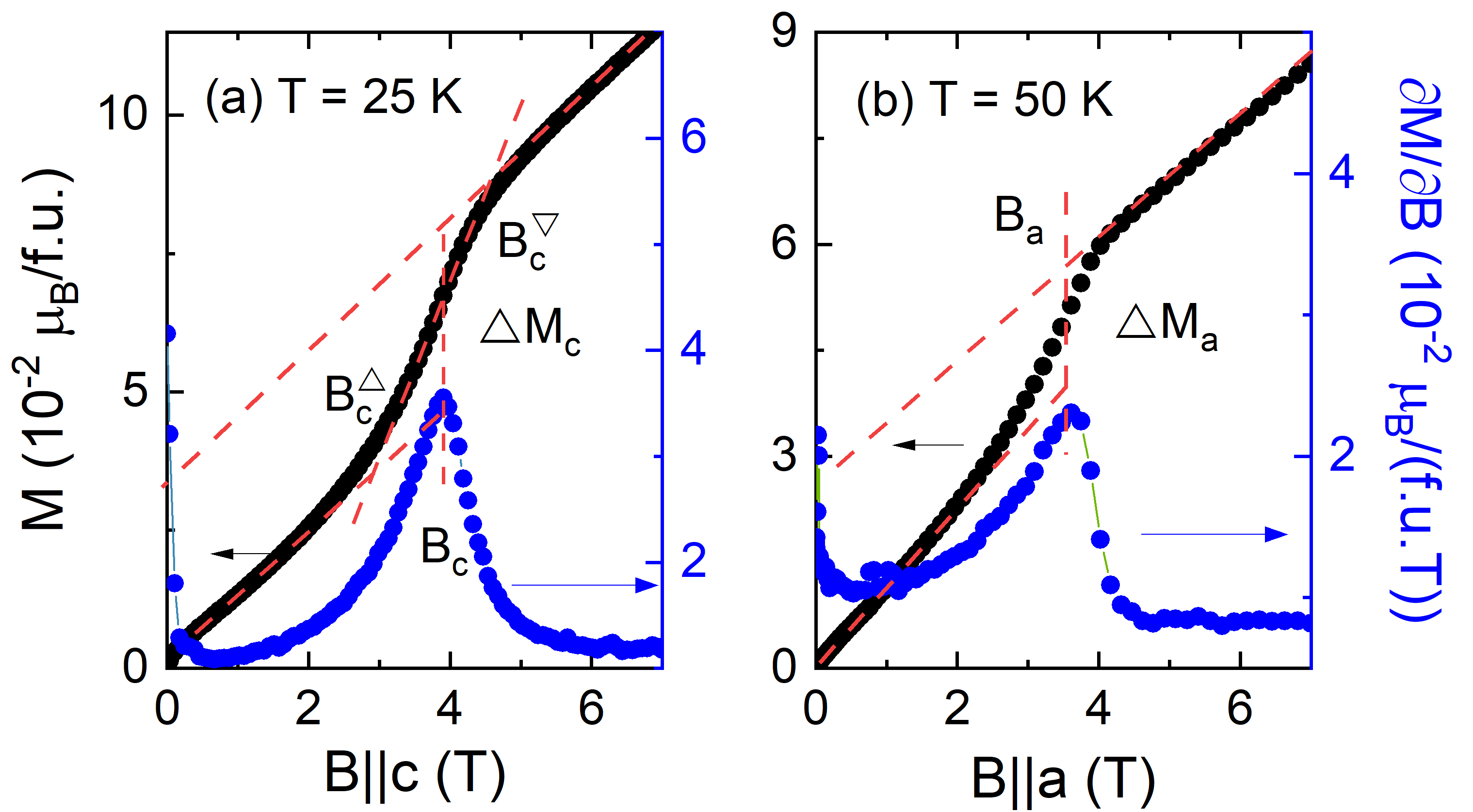}
\caption{Isothermal magnetization and differential magnetic susceptibility for (a) $B||c$, at 25~K ($<$~\tsr ; AF II), and (b) for $B||a$, at 50~K ($>$~\tsr ; AF I). Anomalies are associated with field-induced SRT. $B_{\rm a}$ and $B_{\rm c}$ are the transition fields, and $\Delta M$ is the associated jump in magnetization. }\label{SCO_MB_SRT}
\end{figure} 

\begin{figure}[tb]
\includegraphics[width=0.9\columnwidth,clip]{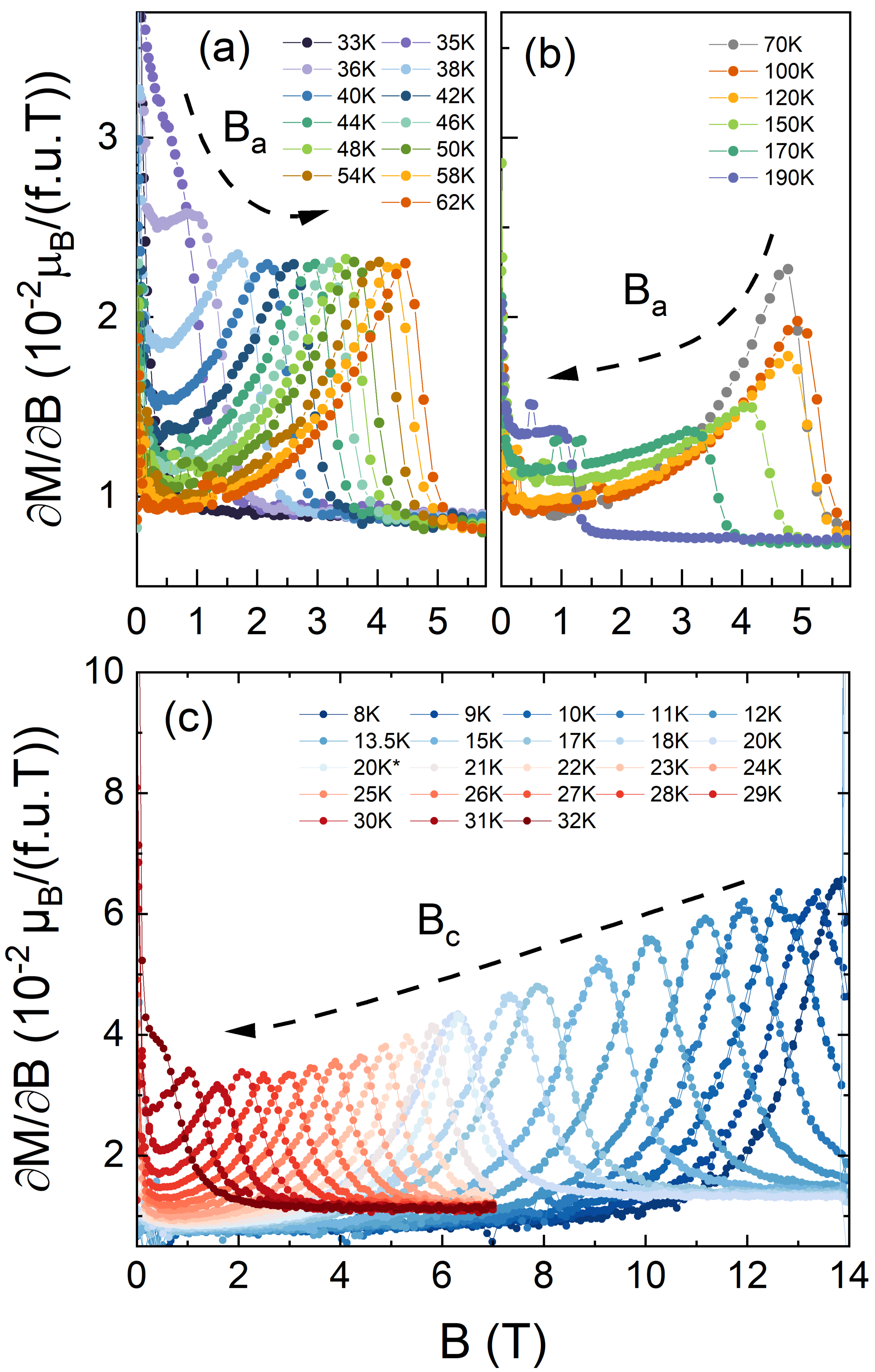}
\caption{Isothermal magnetic susceptibility $\partial M/\partial B$ at various temperatures for (a,b) $B||a$ and (c) $B||c$. The temperature evolution of the SRT is indicated by dashed arrows. $\partial M/\partial B$ measured in the VSM magnetometer ($T\leq 20$~K) have been scaled to the SQUID data ($T\geq 20$~K) such that the anomaly heights at 20~K coincide.}\label{SCO_dMdB}
\end{figure}

Clear anomalies in the field and temperature dependence of the magnetization allow us to construct the magnetic phase diagrams for magnetic fields applied along the different crystallographic axes. This is illustrated for selected temperatures in Fig.~\ref{SCO_MB_SRT} where for $B||a$ and $B||c$, respectively, the magnetization features a jump-like increase as also indicated by the peak in isothermal magnetic susceptibility. At $T=25$~K, magnetic fields $B||c$ induce a small ferromagnetic moment at $B_{\rm c}\simeq 3.9$~T (Fig.~\ref{SCO_MB_SRT}a). For $T>$~\tsr , we do not observe such a feature for $B||c$. We associate the anomaly with field-induced SR towards the $c$ axis, i.e., the transition from AF II to AF I. Whereas, similar features are seen for $B||a$ for \tsr~$\leq T\leq$~\tn\ as illustrated by the example in Fig.~\ref{SCO_MB_SRT}b. This is shown for the whole temperature regime in Fig.~\ref{SCO_dMdB}a,b which shows that $B_{\rm a}$ is increasing upon heating above \tsr\ to a maximum of nearly 5~T, at around 55~K, while again decreasing up \tn . On the other hand, $B_{\rm c}$ continuously increases towards lower temperatures and leaves the accessible field range below 8~K) (Fig.~\ref{SCO_dMdB}c).

\begin{figure}[hbt]
\includegraphics[width=0.9\columnwidth,clip]{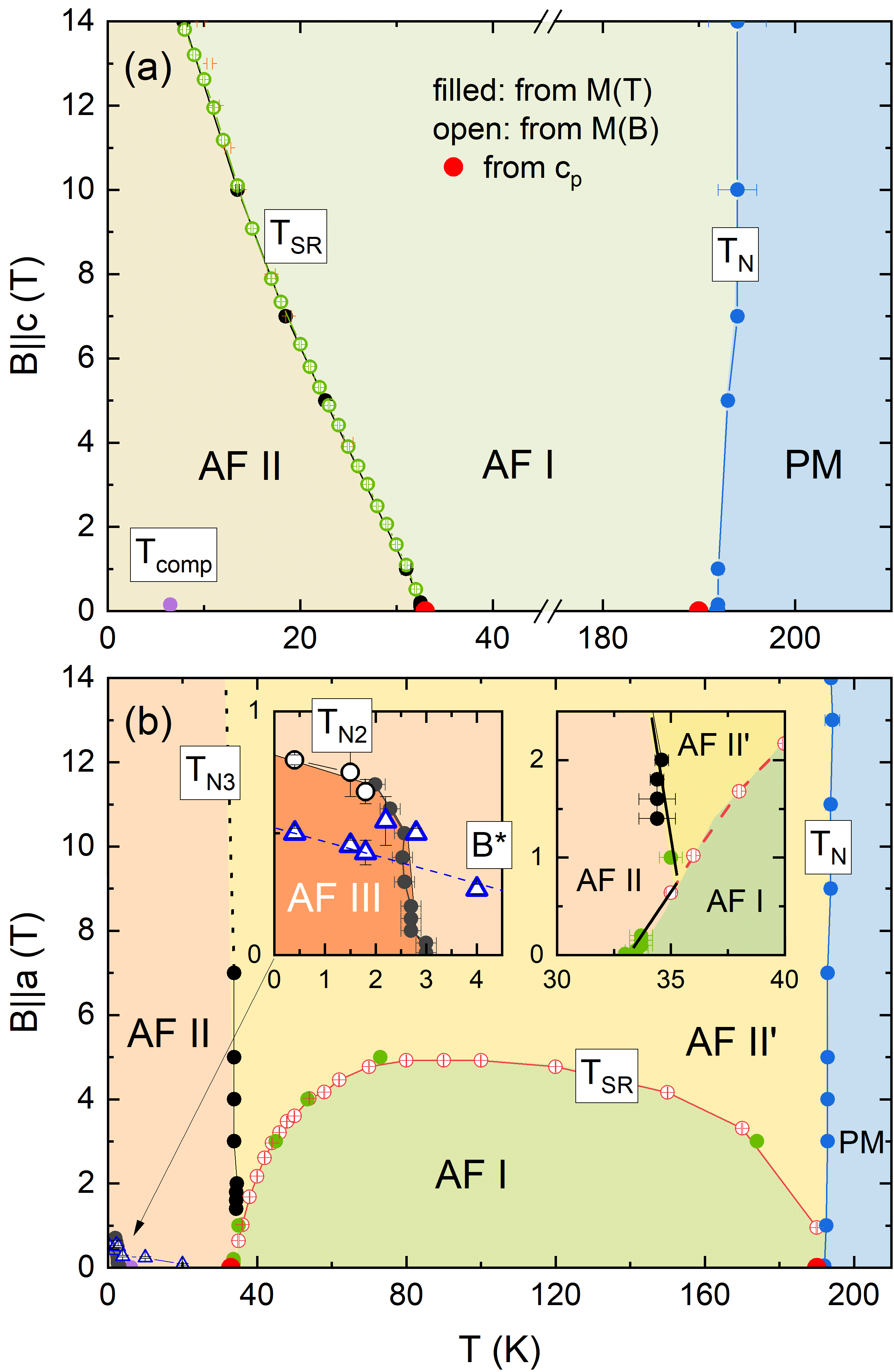}
\caption{Magnetic phase diagram of SmCrO$_3$ for (a) $B||c$ and (b) $B||a$. (See data for $B||b$ in the SM.) Insets in (b) highlight the regimes around 35~K and $T_{\rm N2}(B)$. PM: Paramagnetic phase; AF I: AFM phase with net magnetic moments along the $c$ axis ($\Gamma_4$); AF II and II': AFM phases with net magnetic moments within the $ab$ plane; AF III: AFM phase with ordering of Sm$^{3+}$ moments. Labels of the phase boundaries (\tn , \tsr , \tcomp , \tntwo , \tnthree , $B^*$) are explained in the text.
}\label{SCO_Phase}
\end{figure}

The magnetic phase diagrams for $B||a$ and $B||c$ are constructed from the zero field data in chapter~\ref{ch:srt} and the critical fields, $B_c$ and $B_a$, as shown above. In addition, our magnetization measurements allow us to follow \tn ($B$) and \tnS ($B$) as shown in Fig.~S4a of the SM~\cite{SM}. The resulting phase diagrams are presented in Fig.~\ref{SCO_Phase}. In addition to the features discussed above, we also note a steep phase boundary $T_{\rm N3}(B||a)$ as indicated by anomalies in $M(T, B||a>1.5~\rm{T})$ (see Figs.~\ref{SCO_Phase} and \ref{tri}). For $B||b$, the magnetic phase diagram is similar to what is found for $B||a$ and shown in Fig.~S4 of the SM~\cite{SM}. In the following, we discuss the main findings of the obtained phase diagrams.


For $B||c$, we observe an increase of \tn\ in the whole field range under study. The positive slope $\partial T_{\rm N}/\partial B$ agrees to the fact that the evolution of phase AF I is associated with an increase of magnetization. The absence of pronounced anomalies in magnetic susceptibility for $B\perp c$, on the other hand, agrees to the fact that \tn\ is rather independent on $B||a$ and $B||b$ (see Fig.~\ref{SCO_Phase}b and S4 of the SM~\cite{SM}). Quantitatively, the slope of the continuous phase boundary~\cite{Barron,klingeler2005} is linked to the associated jumps in the heat capacity ($\Delta c_{\rm p}$) and the temperature derivative of the magnetization ($\Delta (\partial M/\partial T)|_B$) as given by the Ehrenfest relation:

\begin{equation}
    \frac{\partial T_{\rm N}}{\partial B}=-T_{\rm N}\frac{\Delta (\partial M/\partial T)|_B}{\Delta c_{\rm p}}.\label{ehrenfest}
\end{equation}

From Fig.~\ref{SCO_MT} we obtain the jump in $\partial M/\partial T$, at $B=0.01$~T of $9(5)\times 10^{-4}$~\mb /f.u. for $B||c$, and $15(3)\times 10^{-2}$~\mb /f.u. for $B||a$. Using
the experimentally determined jump in the specific heat $\Delta c_{\rm p}$ from Fig.~\ref{SCO_Cp0T}, Eq.~\ref{ehrenfest} yields $\partial T_{\rm N}/\partial  B_{||c} = 0.17(13)$~K/T and $\partial T_{\rm N}/\partial  B_{||a} = 3(1)$~mK/T. This agrees to the observed negligible slope $\partial T_{\rm N}/\partial  B_{||a}$ and, within error bars, to $\partial T_{\rm N}/\partial B_{||c}\simeq 0.28(5)$~K/T derived from the phase diagrams in Fig.~\ref{SCO_Phase}.

\subsubsection*{1st order nature of the spin-reorientation transition}

In the literature, there is a controversy about the nature of the SRT in SmCrO$_3$. Based on polycrystal samples, Sau et al.~\cite{sau2022first} reported that the SRT appears in two steps, i.e., a continuous rotation followed by a discontinuous jump, with the former being of second-order and the latter of first-order nature. As shown in Figs.~\ref{SCO_Cp0T} and \ref{SCO_MT}, distinct and single anomalies in our zero field measurements on the single crystal at hand do not support a two-step scenario. The one-step scenario is further corroborated by the following thermodynamic considerations: As shown in Fig.~\ref{SCO_Phase}, magnetic fields $B||c$ yield a monotonous suppression of \tsr , i.e., AF I is stabilised over the spin-rotated low-temperature phase AF II. Microscopically, this is associated with rotation of the weak net magnetic moments of the canted Cr$^{3+}$ sublattice from the $ab$ plane (AF II) to the $c$ axis (AF I). For a discontinuous phase transition, the slope of the phase boundary is governed by the jumps in magnetization ($\Delta M_{\rm{SR}}$) and in the entropy ($\Delta S_{\rm {SR}}$) as described by the Clausius-Clapeyron equation~\cite{Barron}:

\begin{equation}
    \frac{\partial T_{\rm SR}}{\partial B} = - \frac{\Delta M_{\rm SR}}{\Delta S_{\rm {SR}}}. \label{clausius}
\end{equation}

\begin{figure}[htb]
    \includegraphics[width=\columnwidth,clip]{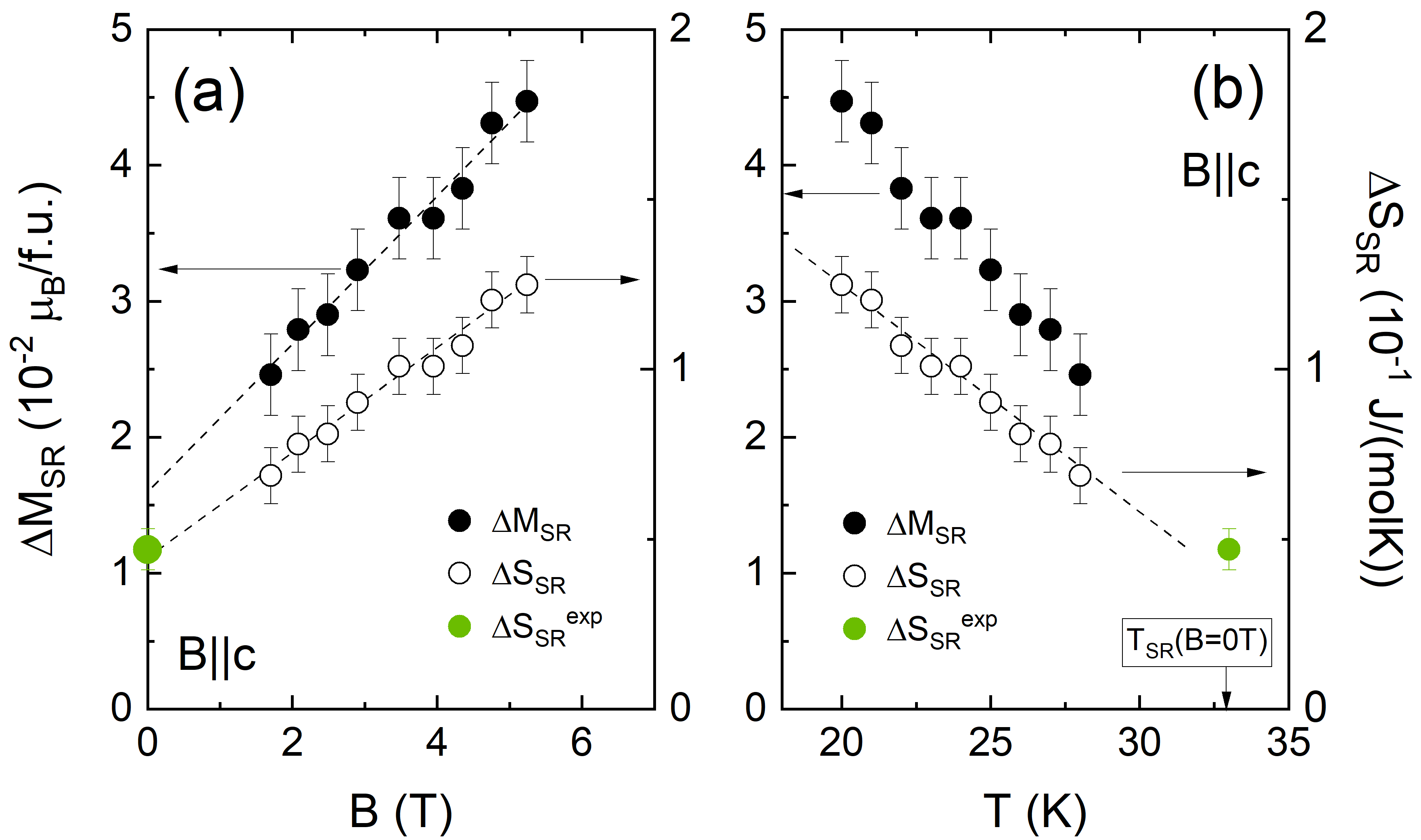}
    \caption{Dependence of magnetization anomaly $\Delta M_{\mathrm {SR}}$ at the SRT (AF I~$\leftrightarrow$~AF~II) extracted from the data as illustrated in Fig.~\ref{SCO_MB_SRT}
    and associated entropy jump $\Delta S_{\mathrm {SR}}$ obtained by means of Eq.~\ref{clausius} on (a) the magnetic field $B||c$ and (b) on the temperature, at different $B||c$. Dashed lines are guides to the eye. Green data markers show $\Delta S_{\mathrm {SR}}$ from the specific heat measurement.}
    \label{SCO_DS}
\end{figure}

The measured slope of the phase boundary $T_{SR}(B||c)$ and magnetization jumps $\Delta M$ hence enable us to calculate the associated entropy changes. The temperature and magnetic field dependence of the experimentally obtained anomaly sizes $\Delta M_{\rm{SR}}$ and the by Eq.~\ref{clausius} calculated entropy jumps are shown in Fig.~\ref{SCO_DS}. Extrapolating the results $\Delta S_{\rm {SR}(B||c)}$ in Fig.~\ref{SCO_DS} to zero field in particular yields $\Delta S_{\rm {SR}} = 0.045(10)$~\jmk\ which nicely agrees to the experimentally measured entropy jump of $\Delta S_{\rm {SR}} = 0.047(1)$~\jmk\ derived from the specific heat data in Fig.~\ref{SCO_Cp0T} (see the green data point in Fig.~\ref{SCO_DS}).

The fact that the entropy jump calculated by the Clausius-Clapeyron equation quantitatively agrees to the experimentally measured discontinuity confirms the measured data and in particular verifies the discontinuous nature of the SRT. We hence exclude the suggested two-step scenario for $B=0$~T. In addition, Fig.~\ref{SCO_DS} shows that upon application of $B||c$ both $\Delta M_{\rm SR}$ and $\Delta S_{\rm SR}$ strongly increase. This implies that magnetic order in AF I is stronger suppressed in favor of the larger magnetization $M_{\rm c}$ than in AF II.

\subsubsection*{Tricritical and triple point for $B||a$}

In contrast to the monotonous and rather constant suppression of \tsr($B||c$), AF I forms a dome in the phase diagram if $B$ is applied along the crystallographic $a$ direction (the same holds for $B||b$, see Fig.~S4 in the SM). As shown in Fig.~\ref{SCO_Phase}b, AF I is stabilised with respect to AF II' for $T\gtrsim 70$~K while being suppressed below and completely vanishes at \tsr($B=0$~T)~=~33~K. However, the behaviour for $B||a$ is rather complex: At temperatures around $\simeq 36$~K, the transition AF II' $\leftrightarrow$ AF I is associated to a jump in $\partial M/\partial B$ and, for $B||a\gtrsim 2$~T, a kink in $M(T)$ (see Figs.~\ref{SCO_dMdB}, S8 and S12) which implies its $continuous$ nature while we recall the discontinuous transition AF~II $\leftrightarrow$ AF I at \tsr ($B=0$~T). The continuous nature of the AF I/AF II' phase boundary is further confirmed by clear failure to describe the slope \tsr ($B||a$) by means of Eq.~\ref{clausius} if anomalies are interpreted as discontinuities. 

\begin{figure}[hbt]
    \includegraphics[width=\columnwidth,clip]{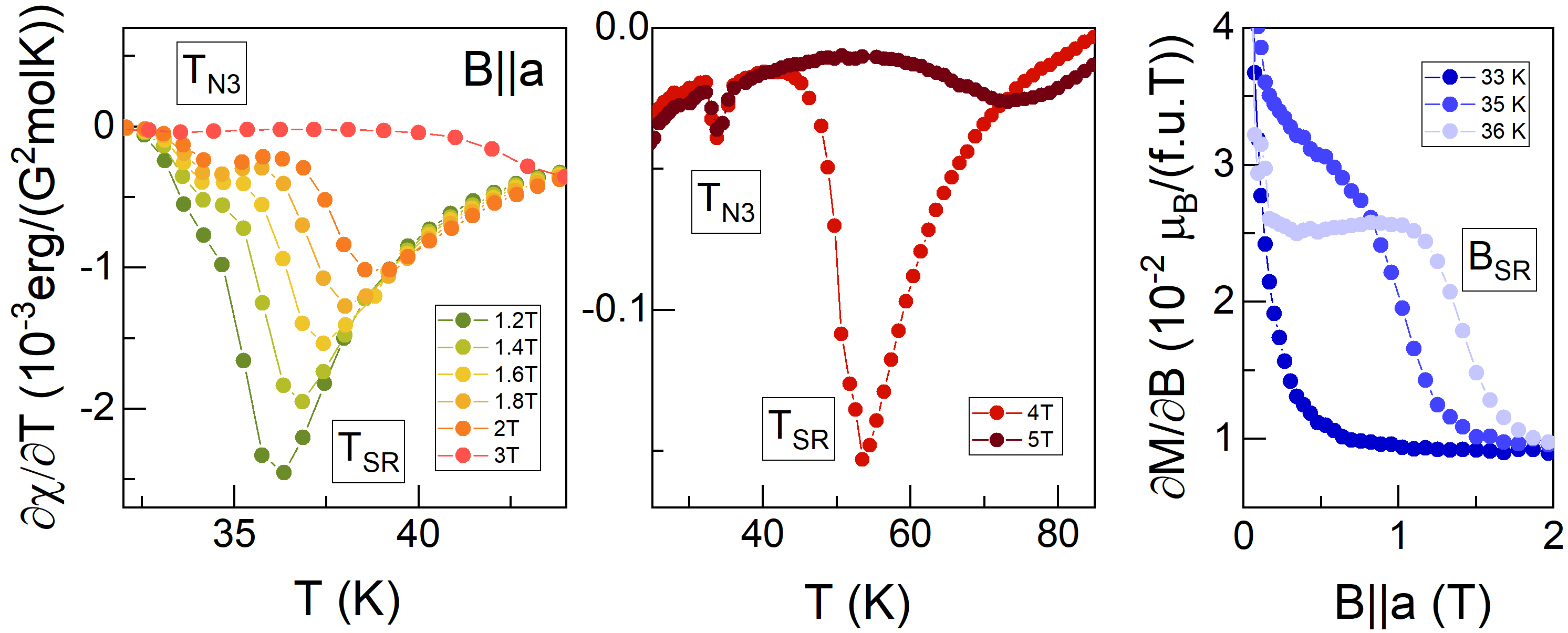}
    \caption{Anomalies in (a,b) $\partial \chi_{\rm a}/\partial T$, and (c) $\partial M/\partial B_{\rm ||a}$ for selected temperatures and fields to illustrate the $B||a$ dependence of \tsr\ and the phase boundaries between AF I, AF II', and AF II. Labels match those in the phase diagram Fig.~\ref{SCO_Phase}. For additional data see the SM~\cite{SM}.
 }
    \label{tri}
\end{figure}

In addition to this continuous-type phase boundary separating AF I and AF II', for $B||a>1.5$~T we observe a small jump-like increase of $M(T)$ which is nearly independent on $B$ (Fig. \ref{tri}b). Our data hence indicate a triple and --- as the jump implies the 1st order character of this AF II/AF II' phase boundary -- a tricritical point at $\simeq 34$~K and $B||a \simeq 1$~T separating phases AF I (moments $||c$), AF II and AF II'. Both latter phases are characterised by moments in the $ab$ plane, with a slightly larger net moment along $a$ in AF II'. 

The transition from AF II' to AF II is associated to finite entropy changes which lower limit can be derived utilizing the anomaly in $M$ and the slope $\partial T_{\rm N3}/\partial B$ using Eq.~\ref{clausius}. For example, the magnetization jump at \tnthree (3~T) is of the order of $\Delta M_{\rm N3}\simeq 3\times 10^{-4}~\mu_{\rm B}$/f.u.. The steep slope of \tnthree ($B||a$) analysed by means of Eq.~\ref{clausius} hence implies $\Delta S_{\rm N3} \gtrsim 20$~mJ/(mol K). 


\subsubsection*{Magnetization compensation and order of Sm$^{3+}$ moments} \label{chap:tntwo}

While the Sm$^{3+}$ sublattice remains paramagnetic (but polarized by the Cr$^{3+}$ sublattice) at \tn , ordering of the 4$f$ moments has been reported at \tntwo~$\simeq$~4~K as deduced from thermodynamic and neutron data~\cite{gorodetsky1977second,gupta2016study,sau2021high}. However, while neutron data at 1.5~K have shown the presence of long-range Sm$^{3+}$ magnetic order, the transition was previously~\cite{gupta2016study} associated with a broad hump in the specific heat similar to what is seen in our data (Fig.~\ref{SCO_Cp0T}). Such a broad hump is however not a typical signature of a thermodynamic phase transition but usually signals the expected Schottky contribution so that, in contrast to Ref.~\cite{gupta2016study}, we do not consider it evidencing \tntwo . Contrarily, we conclude that the evolution of long-range magnetic order of the Sm$^{3+}$ sublattice from the AF II phase (where Sm$^{3+}$ moments are highly polarized) is not associated with significant entropy changes and cannot be detected in the specific heat data against the background of the Schottky peak. In an attempt to estimate an upper limit of the expected specific anomaly by exploiting equation Eq.~\ref{ehrenfest}, we find $\Delta c^{\rm T_{\rm N2}}_{\rm p}\simeq 60$~m\jmk\ which is clearly indistinguishable from the strong Schottky anomaly dominating the total specific heat at \tntwo\ (see Fig.~\ref{SCO_Cp0T}).

\begin{figure}[hbt]
\includegraphics[width=\columnwidth,clip]{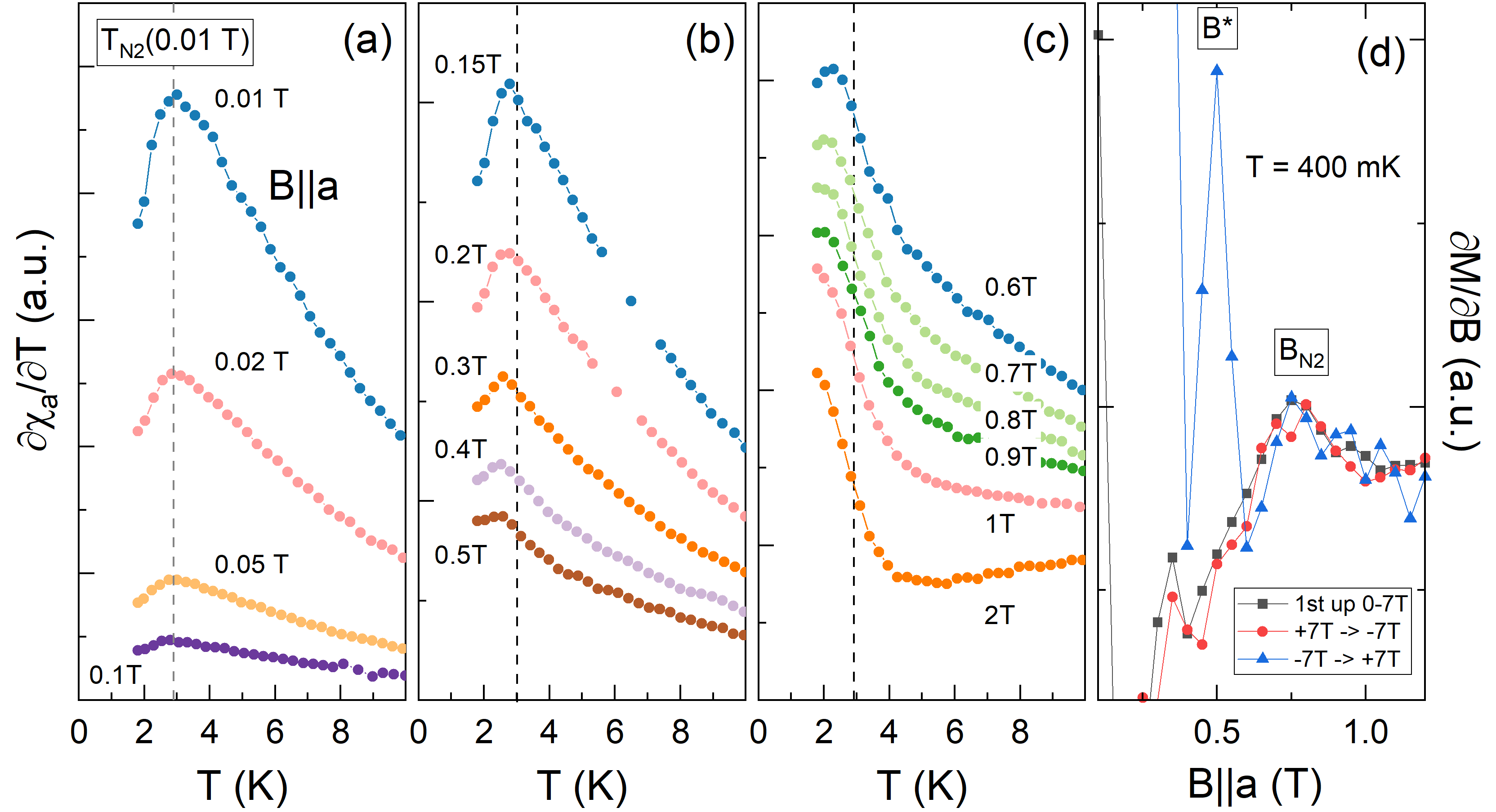}
\caption{Anomalies associated with \tntwo ($B$): (a-c) Temperature derivative of the static magnetic susceptibility, $\partial \chi_{\rm a}/\partial T$ for various fields $B||a$ axis, at low temperatures. Ordinates cover different scales for visibility. The dashed lines show \tntwo\ at 10~mT. (d) Isothermal magnetic susceptibility $\partial M/\partial B$ at $T = 400$~mK for $B||a$ axis. The data illustrate a field sweep 0~T~$\rightarrow$~7~T~$\rightarrow$-7~T~$\rightarrow$~7~T. $B_{\rm N2}$ marks a (non-hysteretic) feature associated with the phase boundary of AF III. The anomaly at $B^*$ only appears after applying high magnetic fields of opposite field direction. For the full sweep see Fig.~S10 in the SM~\cite{SM}.}
\label{SCO_TN2}
\end{figure}

A characteristic feature of long-range Sm$^{3+}$ magnetic order, however, is visible in the temperature dependence of the susceptibility which shows a small kink in $\chi_{\rm a}$ and thus a peak in $\partial \chi_{\rm a}/\partial T$ (see Fig.~\ref{SCO_TN2}a). The kink-like feature confirms the continuous nature of the phase transition. At $B=10$~mT ($||a$), the peak in $\partial \chi_{\rm a}/\partial T$ appears at $T_{\rm N2} \simeq 3$~K which we interpret as the ordering temperature of Sm$^{3+}$ moments. The field dependence of \tnS\ can be followed by the characteristic maximum in $\partial \chi_{\rm a}/\partial T$ as shown in Fig.~\ref{SCO_TN2}a-c. We observe a suppression by $\simeq 1$~K in $B||a = 0.7$~T while for higher fields the peak is not observed (see also Figs.~S16 and S17 of the SM~\cite{SM}). The isothermal magnetization $M(B||a)$ is dominated by the small hysteresis around $B=0$~T. However, we observe small additional peaks as illustrated by the data taken at $T=400$~mK in Figs.~\ref{SCO_TN2}d and S10. Specifically, we observe a tiny hump in $\partial M/\partial B$ both in up and down sweeps which agrees to the phase boundary \tntwo ($B||a$). While this feature (labelled $B_{\rm N2}$ in Fig.~\ref{SCO_TN2}d) vanishes at 2~K, there is a further anomaly ($B^*$ in Fig.~\ref{SCO_TN2}d; see the triangles in the left inset of Fig.~\ref{SCO_Phase}) which appears only after applying high magnetic field in the opposite direction. It is present both in AF III and AF II  (see Fig.~S11 of the SM). The origin of this anomaly in unclear.


The resulting data points down to 400~mK enclosing the low temperature phase AF III are summarized in the inset of the phase diagram in Fig.~\ref{SCO_Phase}b. In the low-temperature phase AF III, both magnetic sublattices develop
long-range magnetic order in agreement to the neutron data from Ref.~\cite{gupta2016study} obtained at 1.5~K in zero magnetic field. The behavior for $B||c$ is less clear: While for $B=10$~mT applied $||c$ we observe a slightly broader peak in $\partial \chi_{\rm c}/\partial T$ at 3~K as well, this feature disappears already at 50~mT so that the phase boundary cannot be distinguished anymore (see Fig.~S17 in the SM).



An additional notable feature of materials with inequivalent sublattices is the potential for a compensation point which appears in SmCrO$_3$ at \tcomp\ = 6~K. It is induced by the antiparallel alignment of {\it R}$^{3+}$ and transition metal ion {\it M}$^{3+}$ magnetic moments, as demonstrated, e.g., in GdCrO$_3$ (\tcomp\ = 144~K, 100~Oe)~\cite{yin2015giant,fita2019spin}, ErFeO$_3$ (\tcomp\ = 46~K, 100~Oe)~\cite{ma2022low,huang2013large}, SmFeO$_3$ (\tcomp\ = 3.9~K, 300~Oe)~\cite{cao2014temperature,marshall2012magnetic},  NdFeO$_3$ (\tcomp\ = 7.6~K, 100~Oe)~\cite{yuan2013spin}, or n $\alpha$-Cr$_3$(PO$_4$)$_2$ (\tcomp\ = 5~K, 1000~Oe)~\cite{Vassiliev}. A theoretical approach by Yamaguchi describes interactions between $R^{3+}$ and Cr$^{3+}$ moments by a temperature-dependent anisotropic effective field acting mainly on the Cr$^{3+}$ moments and following the direction of the net magnetic component of the Cr$^{3+}$ sublattice.~\cite{yamaguchi1974theory} Application of external magnetic field yields a competing energy scale to the temperature-dependent internal field which explains the observed complex field dependencies typically observed in two-sublattice systems such as Dy$_{0.5}$Pr$_{0.5}$FeO$_3$~\cite{wu2014twofold}, YFeO$_3$~\cite{lin2015terahertz}, GdFeO$_3$~\cite{das2017giant}, or ErCrO$_3$~\cite{toyokawa1979spectroscopic}.



\section*{Summary}

We present the high-pressure optical floating-zone growth of SmCrO$_3$ single crystals and the investigation of SmCrO$_3$ magnetic phase diagrams. This study reports, for the first time, 
the experimental parameters for the growth of SmCrO$_3$ single crystals using the floating-zone method. Access to high-quality single crystals enables us to study the magnetic phase diagrams for the different field directions and to clarify the complex magnetism in \sco\ arising from the interplay of anisotropic 3$d$ and 4$f$ magnetic sublattices.

Long-range order of the sublattices appears at \tn\ = 192~K and \tntwo~=~3~K, respectively. In contrast to previous reports on polycrystals, our single crystal data imply a discontinuous and one-step spin-reorientation of net magnetic moments from the $c$ axis into the $ab$ plane at zero magnetic field at \tsr~=~33~K. Its discontinuous nature is maintained if $B$ is applied $||c$ axis. Our analysis indicates that the spin configuration is $\Gamma_4$ for $T > T_{\mathrm {SR}}$, while it is dominated by $\Gamma_2$ for $T < T_{\mathrm {SR}}$. Our single crystal study in particular rules out that the collinear antiferromagnetic structure $\Gamma_1$ may be realised in AF II below 10~K as reported in Ref.~\cite{tripathi2017evolution} but confirms a considerable uncompensated moment in the spin-reoriented phases AF II and AF III. While the presence of a net magnetic moment in principle is consistent with the $\Gamma_2$ configuration as suggested in Ref.~\cite{sau2021high}, we find clear deviations from the $\Gamma_2$ behavior. 
In addition, while the size of the net magnetic moment decreases upon cooling below $\sim 20$~K, the interplay of the two magnetic sublattices also results in a temperature variation of the in-plane net magnetic moment at \tntwo\ when long-range Sm$^{3+}$ magnetic order evolves and the AF III phase is established. \rk{For AF III, our data suggest but do not unambiguously prove the presence of a small remaining moment $||c$; the observation that the in-plane angle $\alpha$ is still finite however suggests that the $\Gamma_2$ configuration not realized in AF III either.}

When applying finite magnetic fields, notably, we find a triple point as well as tricritical behavior for $B||a$ axis which further highlights the complex interplay of anisotropic magnetic moments in \sco . The magnetic phase diagrams of SmCrO$_3$ for fields along all crystallographic directions down to 0.4~K and up to 14~T provide a crucial foundation for future detailed investigations of the complex magnetic interactions and magnetoelectric coupling in orthochromites.

\begin{acknowledgments}

Support by Deutsche  Forschungsgemeinschaft (DFG) under Germany’s Excellence Strategy EXC2181/1-390900948 (The Heidelberg STRUCTURES Excellence Cluster) is gratefully acknowledged. N.Y. acknowledges fellowship by the Chinese Scholarship Council (File No. 201906890005). L.B. acknowledges funding by the German Research Foundation (DFG) via Research Training Group GRK 2948/1. The authors thank Ilse Glass, Siegmar Roth, and Andre Beck for technical support of x-ray diffraction.

\end{acknowledgments}

The Supplemental Material contains further information on the grown crystal, x-ray diffraction, and magnetization data as well as the phase diagram for fields $\|b$ axis, including references~\cite{white1969review,sco-xrd2017evolution}.



\bibliography{SCO}

\newpage

\section{Supplemental Material}

\section*{Pictures of growth chamber, obtained single crystal, and Laue pattern}

\begin{figure}[h]
\centering
\includegraphics [width=\columnwidth,clip] {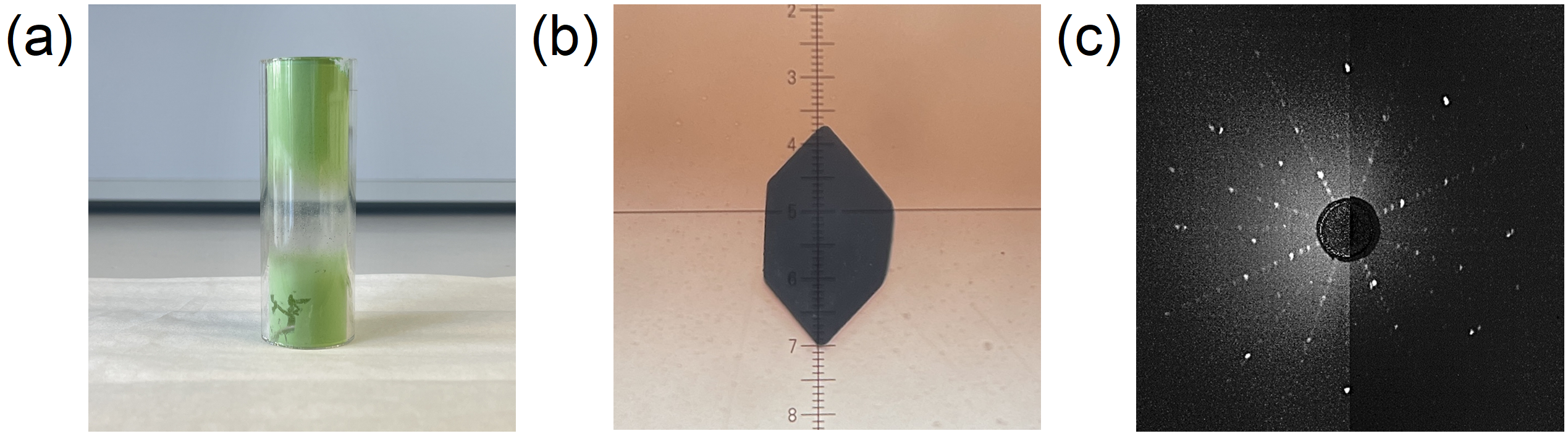}
\caption{(a) Picture of the protective tube using during the high-pressure optical growth process which shows volatiles Cr$_2$O$_3$ attached to the inner wall. SmCrO$_3$ undergoes decomposition during growth, leading to the loss of Cr$_2$O$_3$ due to volatilization. This loss results in an imbalance of the stoichiometric ratio, and the volatiles adhering to the inner wall of the protection tube affect the focusing of the light source. (b) The obtained oriented single crystal. (c) Laue diffraction pattern of the SmCrO$_3$ single crystal oriented along the [001] direction.} \label{FigS1}
\end{figure}

\begin{figure}[hbt]
\includegraphics[width=0.8\columnwidth,clip]{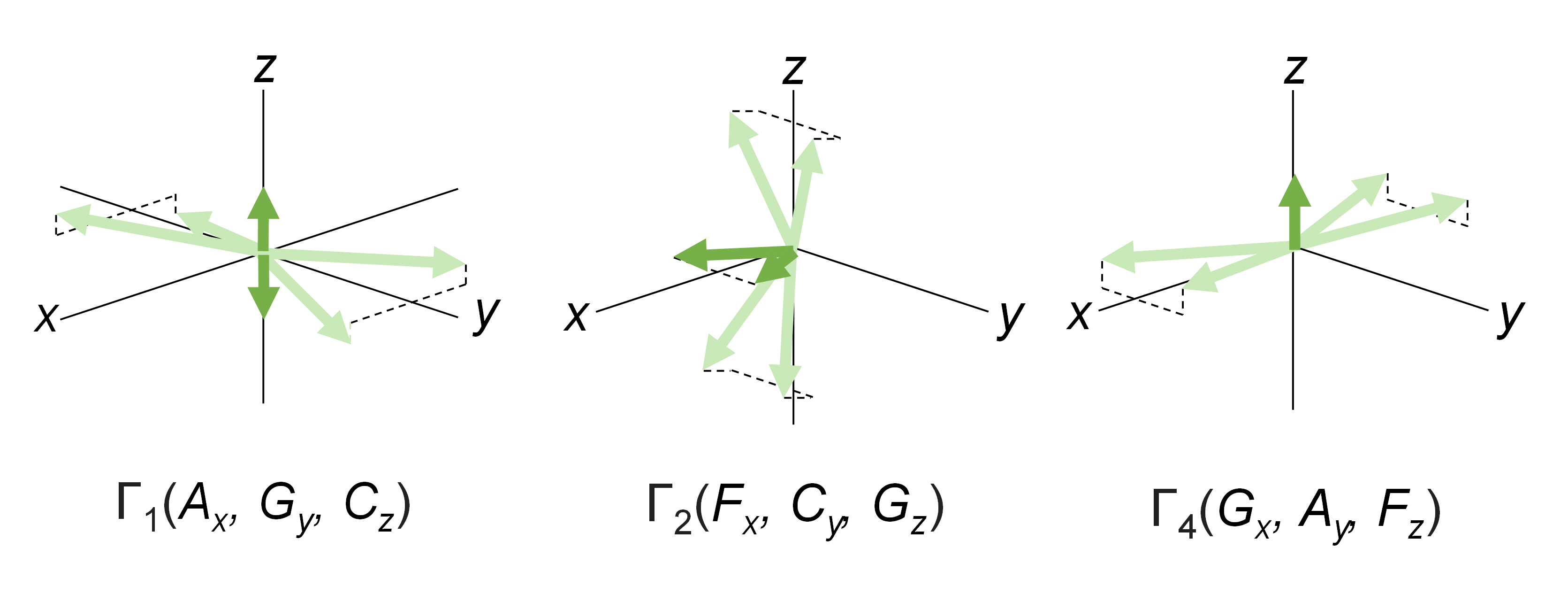}
\caption{Schematics of possible spin configurations in $R$CrO$_3$~\cite{white1969review,hornreich1978magnetic,yamaguchi1974theory}. The light and dark green arrows depict the Cr$^{3+}$ and net moments, respectively. $R^{3+}$ moments are not shown. After \cite{white1969review}.}\label{sketch}
\end{figure}

\newpage

\section*{Structure parameters}

\begin{table}[htb]

\caption{\label{tab:table2}
Refined structural parameters for \sco\ ($Pbnm$) at room temperature. The refinement is performed based on the ICSD No.~5988~\cite{sco-xrd2017evolution}} 

\begin{ruledtabular}
\begin{tabular}{ccccccc}
Atoms &Wyckoff position &\textit{x} &\textit{y} &\textit{z} &Lattice Parameters(\AA) &Reliability factors \\ \hline
Sm &4$c$ &-0.01030 &0.05048 &0.25000 &$a$= 5.3646(1) &\rwp\ = 13.3$\%$                                                                        \\
Cr &4$b$ &0.50000 &0.00000 &0.00000 &$b$= 5.5025(4) & $\chi^{2}$ = 1.78                                                                       \\
O1 &4$c$ &0.08780 &0.47970 &0.25000 &$c$= 7.6437(4) &                                                                        \\
O2 &8$d$ &-0.29000 &0.28510 &0.04340 & $\alpha = \beta = \gamma = 90^{\circ}$&                                                                                \\
                                                
\end{tabular}
\end{ruledtabular}
\end{table}


\begin{table}[h!]
\caption{Crystallographic results of SmCrO$_3$ as determined from single-crystal x-ray diffraction at 300 K and 80 K. The structure was refined in the orthorhombic space groups $Pbnm$ and $Pbn2_{1}$ 
($\alpha = \beta = \gamma = 90^{\circ}$). The lattice parameters, $a$, $b$, and $c$ are shown together with the Wyckoff positions of the atoms, and the equivalent atomic displacement parameters $U_{eq}$. The ADPs were refined anisotropically but due to space limitations only the  $U_{eq}$ are listed in the Table.\@ $f$ is the Flack parameter (see text). Errors shown are statistical errors from the refinement.}

\label{tab:Crystallographic data_300K}

\begin{ruledtabular}

    \begin{tabular} {c c c c c c}

       &  &\textbf{300 K}& \textbf{300 K}&\textbf{80 K}& \textbf{80 K}\\ [0.5ex]  
        \hline\hline
       & SG &  $Pbnm$ &  $Pbn2_{1}$& $Pbnm$ & $Pbn2_{1}$\\   
        & $a$(Å)&  5.3657(1) & 5.3657(1) &  5.3583(1)& 5.3583(1)
\\  
 & $b$(Å)&  5.5072(1)&  5.5072(1)& 5.4988(1) & 5.4988(1)
        
\\  
          & $c$(Å)& 7.6489(1) &  7.6489(1)& 7.6340(1) & 7.6340(1)
\\   
        \textbf{Sm} &  Wyck.&  4c&  4a& 4c&  4a
\\   
         &  $x$  & 0.511248(15) &  0.511221(14)& 0.511099(13) &0.511076(12)   

\\

&  $y$   &  0.948740(17)&  0.948753(15)& 0.948316(14)  &  0.948330(12)     
\\
&  $z$   & 3/4 &  0.75008(4)& 3/4 & 0.75014(3) 
\\

&  $U_{eq}$&  0.00453(2)&   0.004591(19)& 0.001780(19)& 0.001843(17) 
\\

        \textbf{Cr} &  Wyck.&  4a&  4a& 4a&  4a
\\   
         &  $x$    &  0&  0.0015(3)& 0& 0.0007(3)  \\ 

&  $y$  &  0&  0.0004(4)& 0 & 0.0009(4)             
\\
&  $z$   &  0&  0.0005(2)& 0& 0.0004(1) 
\\

&  $U_{eq}$&  0.00308(5)&  0.00310(5)&  0.00169(5)& 0.00173(4) 
\\

        \textbf{O1} &  Wyck.&  4c&  4a& 4c&  4a
\\   
         &  $x$  & 0.0876(3) &  0.0876(2)&   0.0873(2) & 0.0871(2)   \\ 

&  $y$   &  0.0240(2)&  0.0242(2)& 0.0239(2) & 0.0240(2)              
\\
&  $z$    &  3/4 &  0.7495(11)& 3/4 & 0.7492(8) 
\\

&  $U_{eq}$&  0.0057(2) &  0.0057(2)& 0.0039(2)& 0.0040(2) 
\\

        \textbf{O2} &  Wyck.&  8d&  4a& 8d&  4a
\\   
         &  $x$   & 0.7028(2) &  0.6980(13)& 0.7026(1) & 0.6980(11)       \\ 

&  $y$    &  0.20477(16)&  0.2100(7)& 0.20479(15) & 0.2080(7)                  
\\
&  $z$     & 0.95419(11) &  0.9496(7)& 0.95446(10) & 0.9520(6) 
\\

& $U_{eq}$&  0.00560(17) &  0.0053(7)& 0.00346(14)& 0.0049(7) 
\\
        \textbf{O3} &  Wyck.&  -&  4a& -&  4a
\\   
         &  $x$    &  -&  0.2927(13)& - & 0.2930(11)       \\ 

&  $y$   &  -&  0.8001(6)& - & 0.7978(6)                     
\\
&  $z$     &  -&  0.0419(7)& - & 0.0437(6) 
\\

&  $U_{eq}$&  - &  0.0043(6)& -& 0.0016(6)
\\
 & 	\# parameters & 29&  48& 29& 48\\
 & 	$f$ & - &  0.516133& - & 0.505201\\
& wR$_2$ (\%) &  3.79&  4.30&  3.51& 3.84 \\
 
& R$_1$ (\%) &  2.07&  2.45&  1.68& 1.96 \\
& GOF &  1.09& 1.07&  1.09& 1.03 
         
\\ [1ex]

    \end{tabular}
\end{ruledtabular}
\end{table}

\pagebreak

\section*{Moriya Model}

\begin{figure}[h]
\centering
\includegraphics [width=0.8\columnwidth,clip] {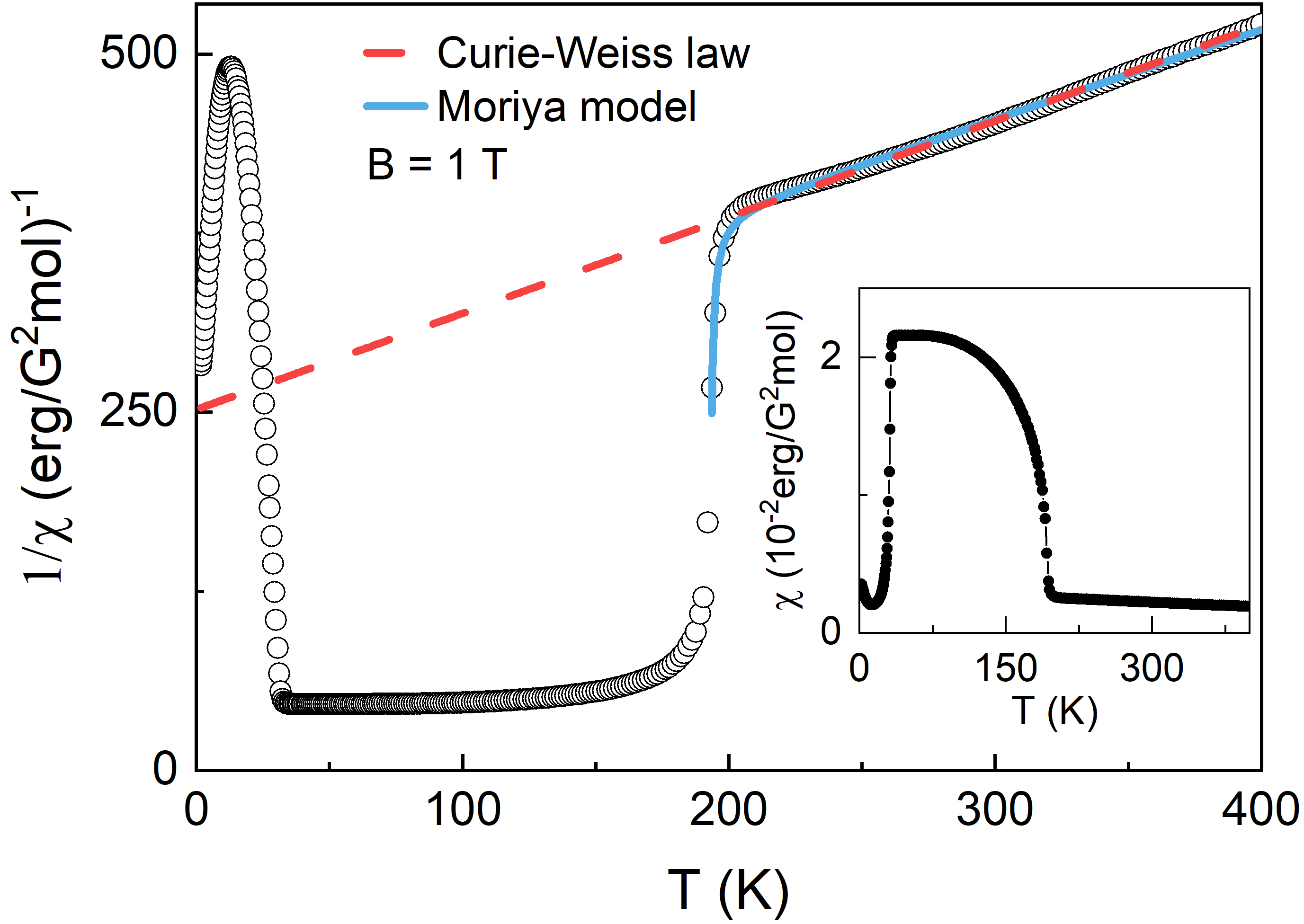}
\caption{Inverse static susceptibility $\chi^{-1} = (M/B)^{-1}$ (inset: $\chi$), obtained at $B = 1$~T applied for $B||c$ of SmCrO$_3$ fitted by the Curie-Weiss law and Moriya model~\cite{moriya1960anisotropic}.} \label{SCO_cwFit}
\end{figure}

As suggested by Moriya, the basic Curie-Weiss (CW) description of the high-temperature static magnetic susceptibility can be extended by incorporating the antisymmetric exchange interaction (Dzyaloshinskii-Moriya (DM) interaction)~\cite{moriya1960anisotropic}. Moriya's model states that when magnetic fields are applied along the magnetic easy axis, the susceptibility adheres to the CW law. Nevertheless, for fields applied perpendicular to the easy axis, the susceptibility can be described by the following Eq.~\ref{cw_moriya}:

\begin{equation}
    \chi = \frac{N_{\mathrm A}\mu_{\mathrm{eff}}^{2}}{3 k_{\mathrm B}(T-\Theta_{\mathrm{W}})}\frac{(T-T_{\mathrm 0})}{(T-T{_\mathrm N}^{\mathrm {Cr}})}, 
    \label{cw_moriya}
\end{equation}

\begin{equation}
  \mathrm{with}~T{_\mathrm 0} = \frac{2J{_\mathrm e}ZS(S+1)}{3 k_{\mathrm B}}~\mathrm{, and}~T_{\mathrm N}^{\mathrm {Cr}} = \frac{2J{_\mathrm e}ZS(S+1)}{3 k_{\mathrm B}}[1+(\frac{D}{2J})^{2}]^{1/2}.
    \label{cw_moriya_T0}
\end{equation}


$N_{\mathrm A}$ is Avogadro's number, $k_{\mathrm B}$ the Boltzman constant, and $\Theta_{\mathrm W}$ the Weiss temperature. Fitted parameters are $T_{\mathrm 0}$, the N$\acute{e}$el temperature for magnetic ordering of the Cr sublattice $T_{\mathrm N}^{\mathrm {Cr}}$, as well as the symmetric (antisymmetric) exchange interactions between Cr$^{3+}$ ions, $J$ ($D$). $S=3/2$ is the spin quantum number of Cr$^{3+}$, and $Z = 6$ represents the coordination number of Cr$^{3+}$ concerning other Cr$^{3+}$~\cite{mcdannald2015magnetic,yin2017magnetic}. 

\newpage 

\section*{Magnetisation and data analysis}

\begin{figure}[hbt]
\includegraphics[width=\columnwidth,clip]{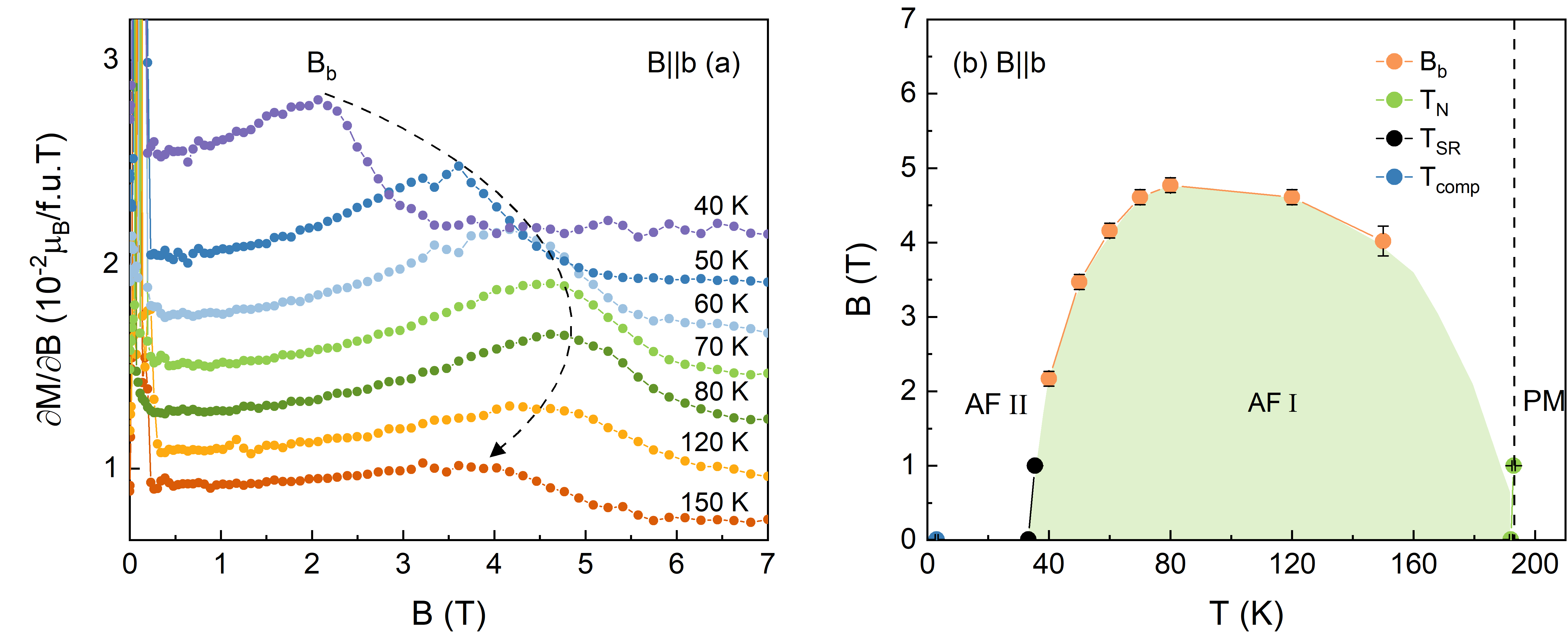}
\caption{(a) Magnetic susceptibility $\partial M/\partial B_{||b}$ for $B||b$ of SmCrO$_3$ at different temperatures. The dashed line indicates the evolution of $B_{\mathrm b}$. The curves are offset vertically by 2.2$\times$10$^{-3}$~$\mu_{\mathrm B}/{\mathrm {f.u.}}$ for better visibility. (b) Magnetic phase diagram of $B||b$ for SmCrO$_3$. PM: Paramagnetic phase; AF I: Antiferromagnetic phase with net magnetic moments are along the $c$ axis ($\Gamma_4$); AF II: Antiferromagnetic phase with net magnetic moments ordered in the $ab$ plane.}\label{SCO_MBb}
\end{figure} 

\begin{figure}[hbt]
\includegraphics[width=\columnwidth,clip]{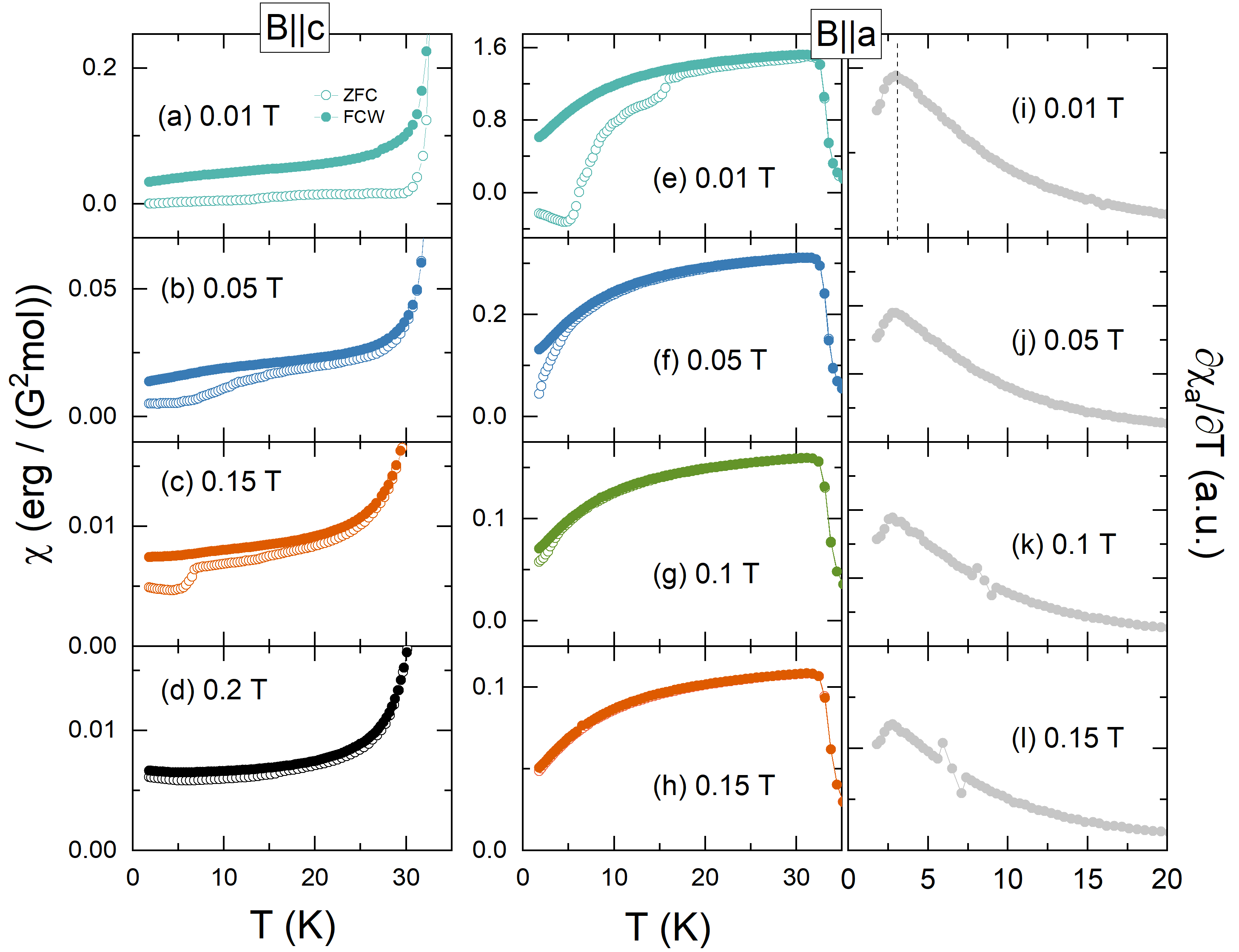}
\caption{Temperature dependence of the static magnetic susceptibility $\chi = M/B$ at different fields $B||c$ axis (a-d) and $B||a$ axis (e-h). ZFC (FCW) data are shown by open (solid) circles. (i-l) show $\partial \chi_{\rm a}/\partial T$ (right ordinate). The vertical dashed lines is a guide to the eye.} \label{SM_chi_fc-zfc-derivative}
\end{figure}

\begin{figure}[hbt]
\includegraphics[width=0.7\columnwidth,clip]{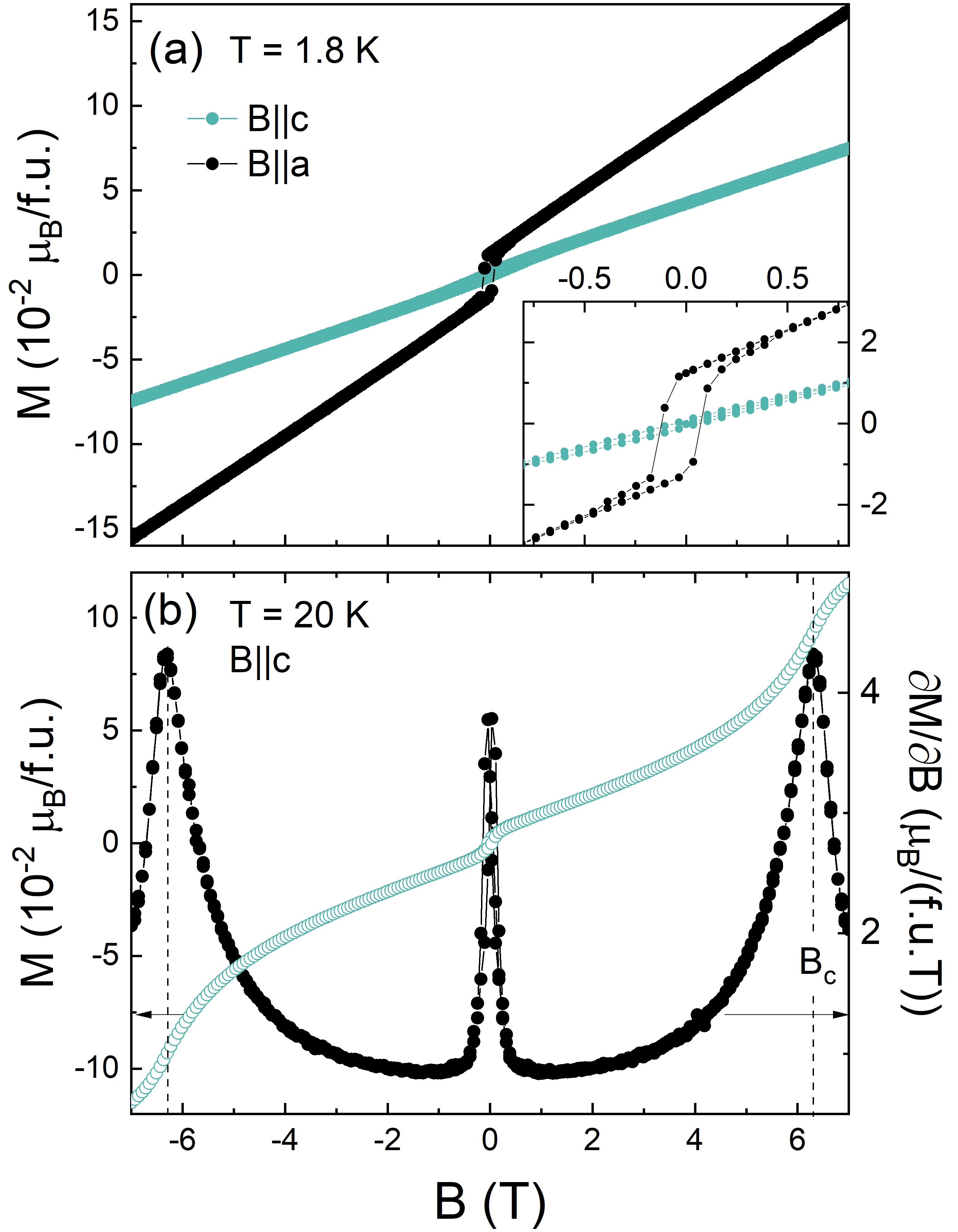}
\caption{Isothermal magnetization (a) at 1.8~K, for $B||a$ and $B||c$, and (b) at 20~K, for $B||c$, with magnetic susceptibility $\partial M/\partial B$. The inset in (a) highlights the behaviour around $B=0$~T.}\label{SM_MBfull}
\end{figure} 

\begin{figure}[tb]
\includegraphics[width=\columnwidth,clip]{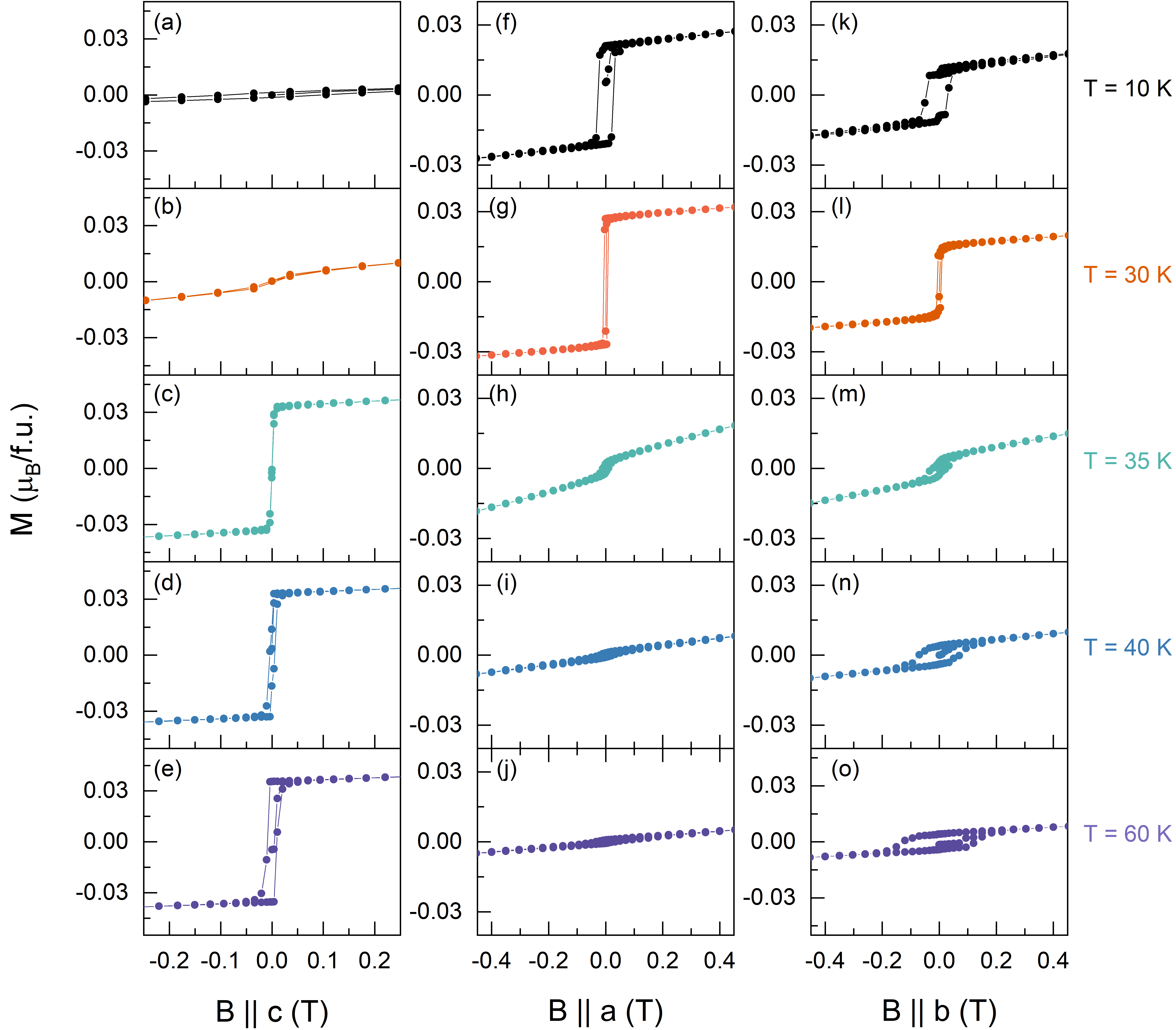}
\caption{Isothermal magnetization at selected temperatures for $B$ applied along the three crystallographic axes.}\label{SCO_MBall}
\end{figure}

\begin{figure}[hbt]
\includegraphics[width=0.8\columnwidth,clip]{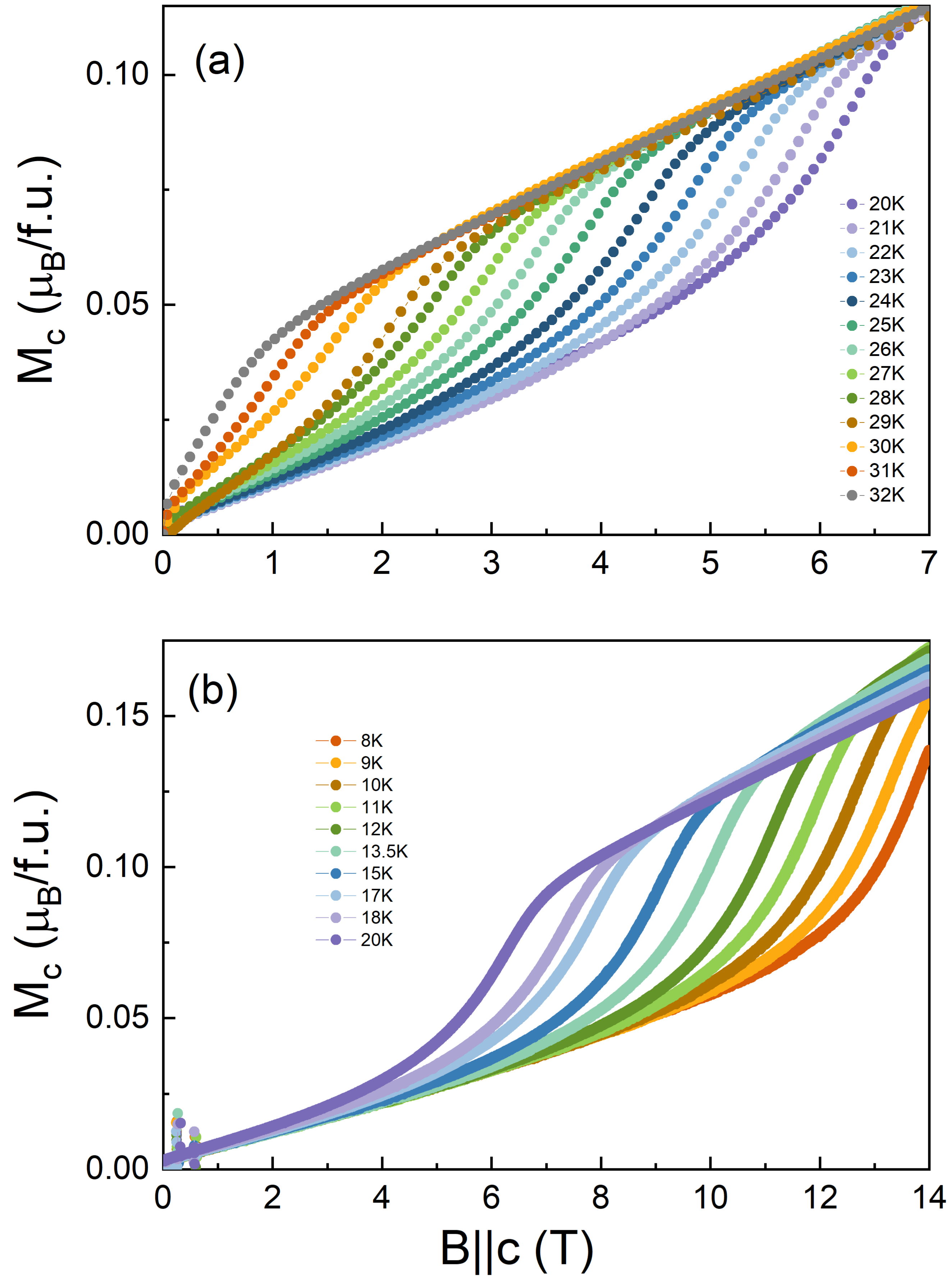}
\caption{Isothermal magnetization for $B||c$ axis at various temperatures. The derivatives of the data are shown in the main manuscript file.}\label{SM_MBc}
\end{figure} 

\begin{figure}[hbt]
\includegraphics[width=0.8\columnwidth,clip]{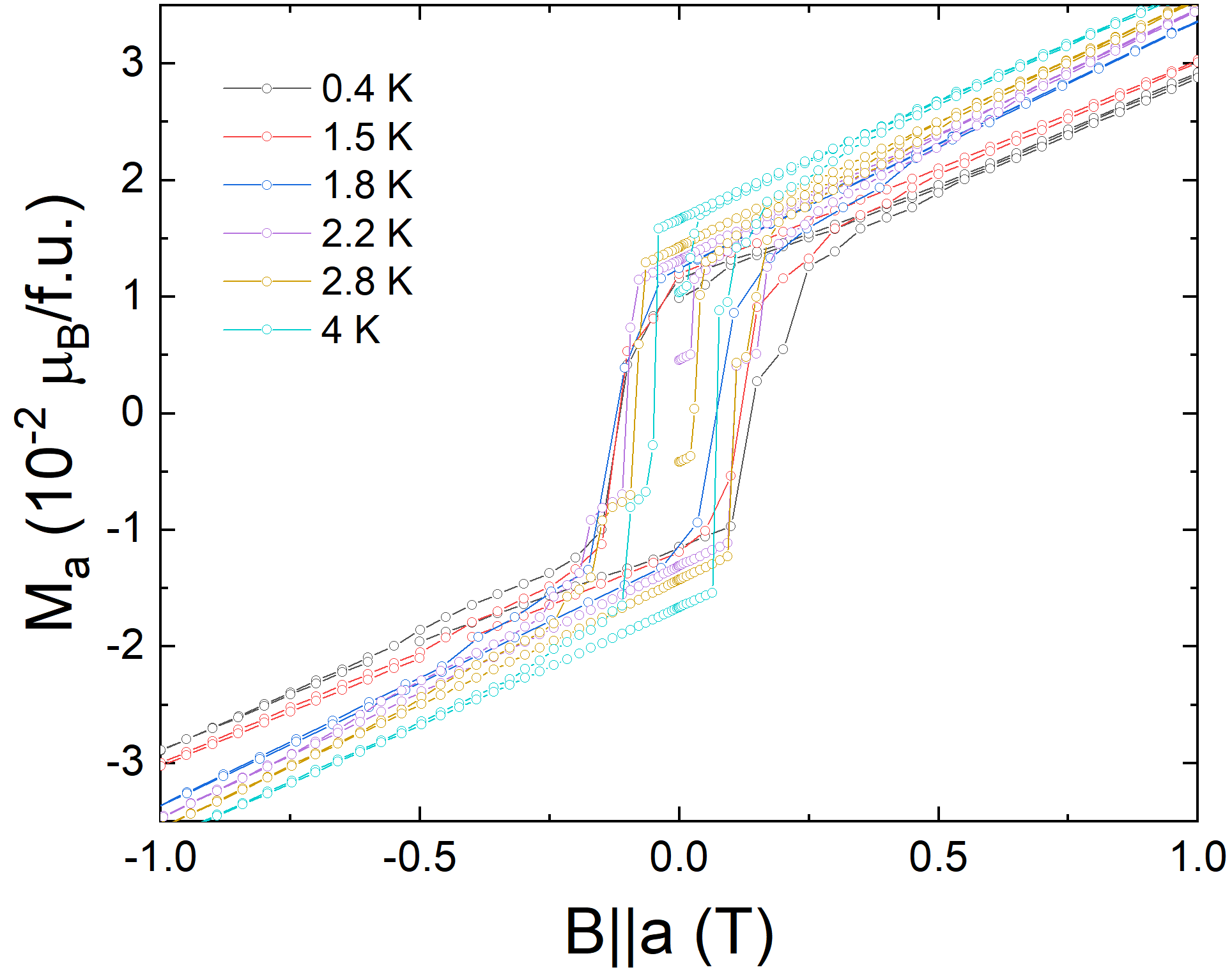}
\caption{Isothermal magnetization for $B||a$ axis at various temperatures}
\end{figure}

\begin{figure}[hbt]
\includegraphics[width=0.8\columnwidth,clip]{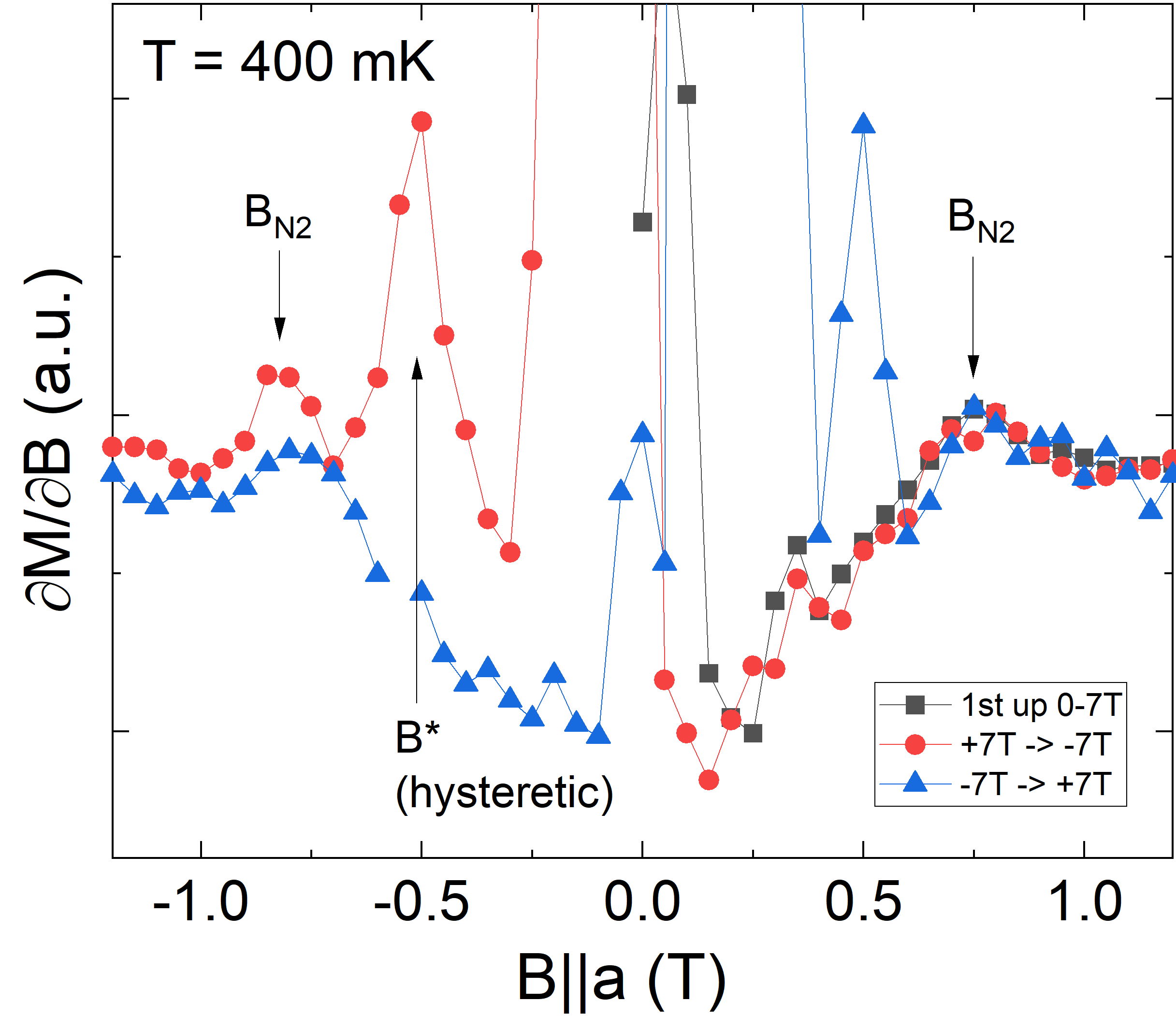}
\caption{Isothermal magnetic susceptibility $\partial M/\partial B$ at $T = 400$~mK for $B||a$ axis. The data illustrate a field sweep 0~T~$\rightarrow$~7~T~$\rightarrow$-7~T~$\rightarrow$~7~T. $B_{\rm N2}$ marks a (non-hysteretic) feature associated with the phase boundary of AF III. The anomaly at $B^*$ only appears after applying high magnetic fields of oppositite field direction.}   
\label{SCO_400mK}
\end{figure} 

\begin{figure}[hbt]
\includegraphics[width=\columnwidth,clip]{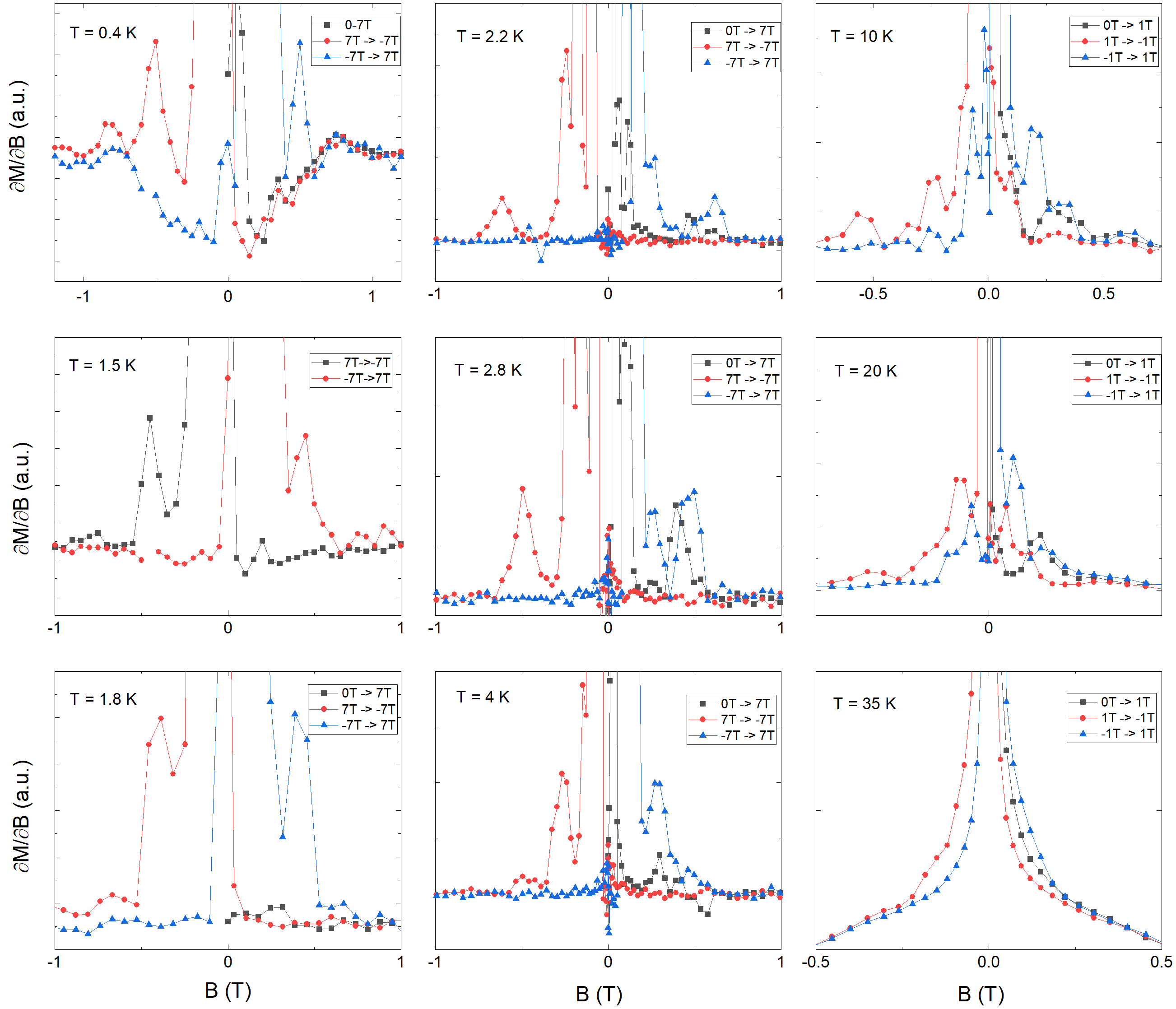}
\caption{Derivative of the isothermal magnetization for $B||a$ axis at various temperatures.}
\end{figure}

\begin{figure}[hbt]
\includegraphics[width=0.7\columnwidth,clip]{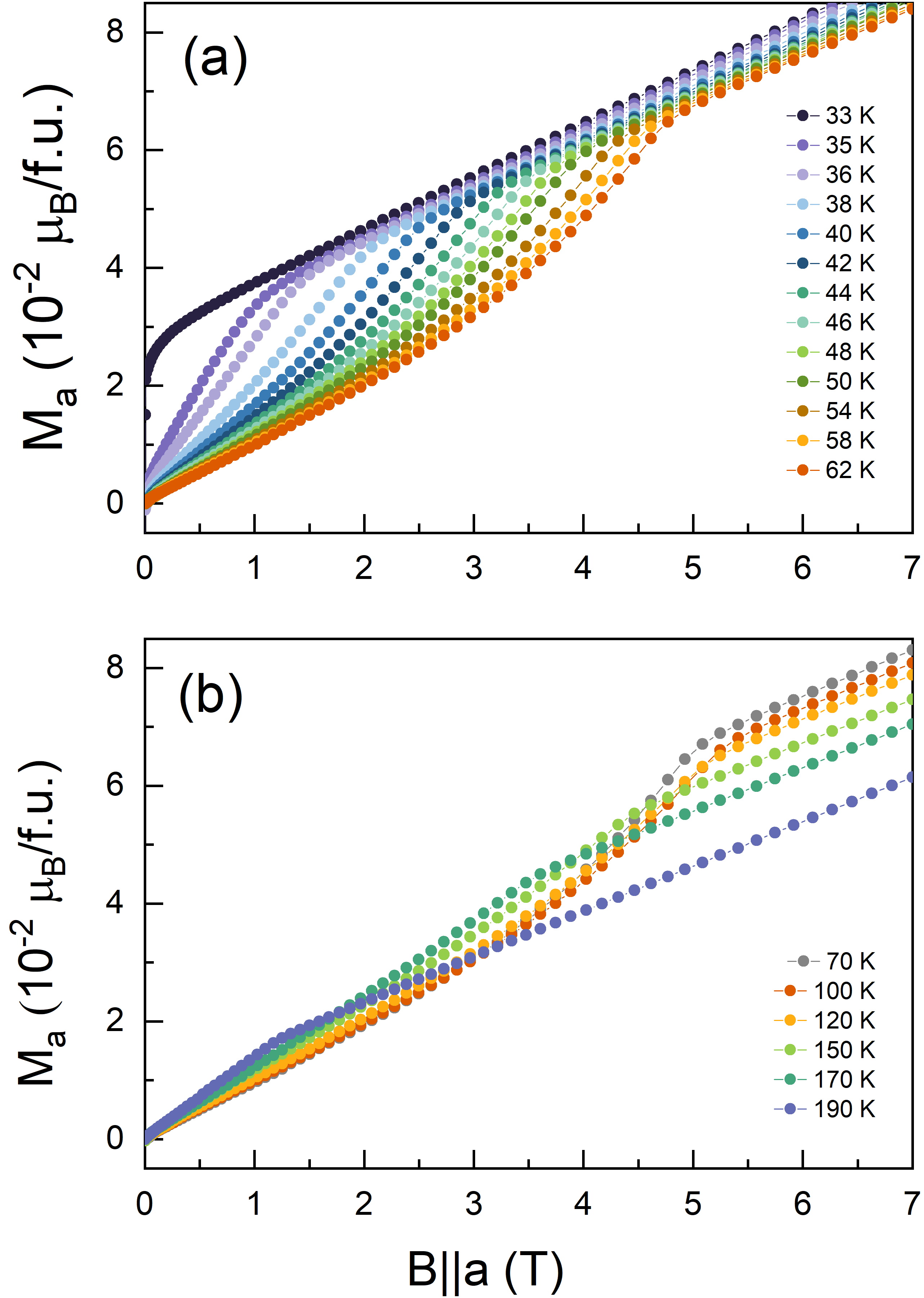}
\caption{Isothermal magnetization for $B||a$ axis at various temperatures. The derivatives of the data are shown in the main manuscript file.}\label{SM_MBa}
\end{figure}

\begin{figure}[hbt]
\includegraphics[width=\columnwidth,clip]{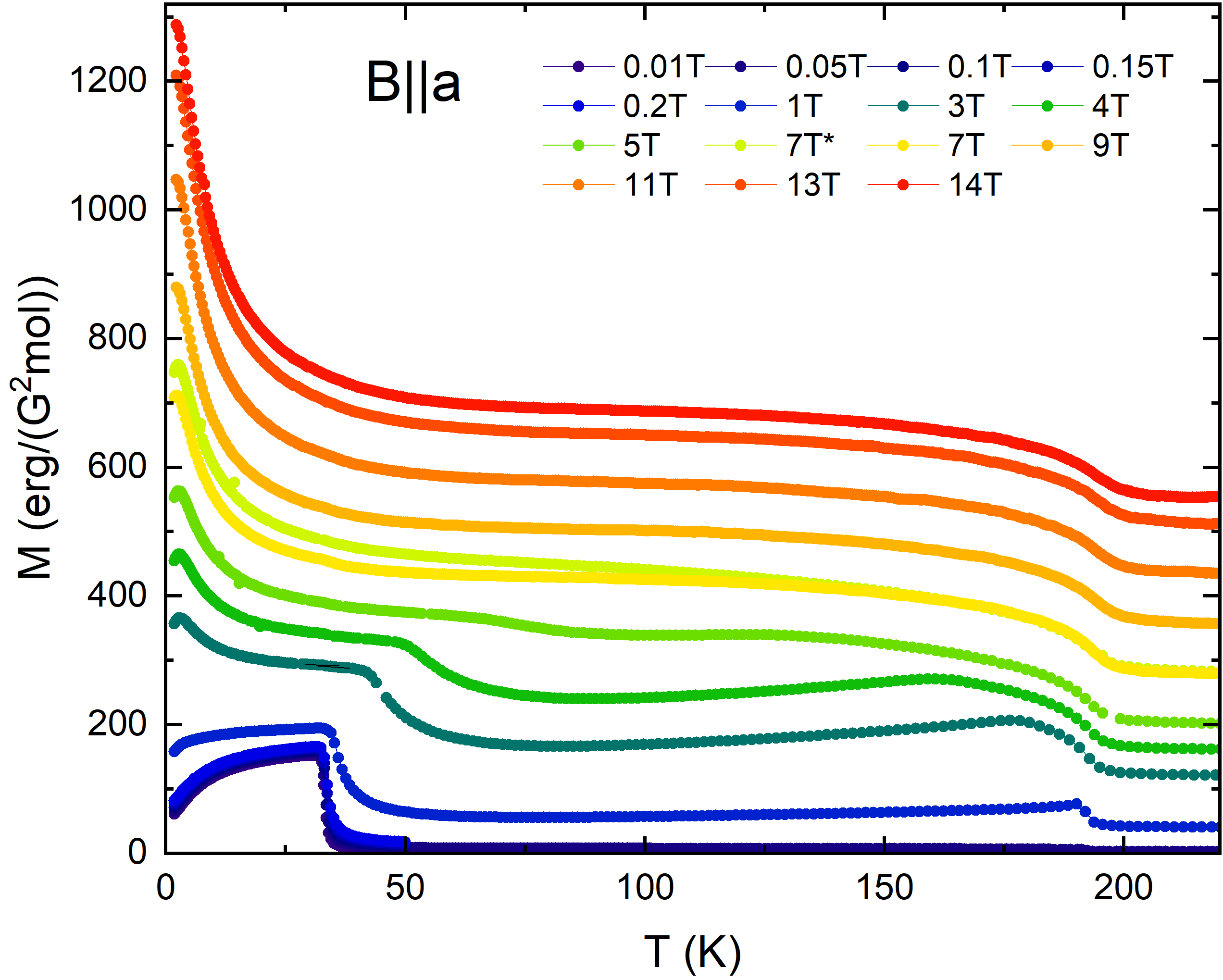}
\caption{Magnetization vs. temperature of \sco\ at different magnetic fields $B||a$ axis. }\label{SM_Ma_overview}
\end{figure}

\begin{figure}[hbt]
\includegraphics[width=0.8\columnwidth,clip]{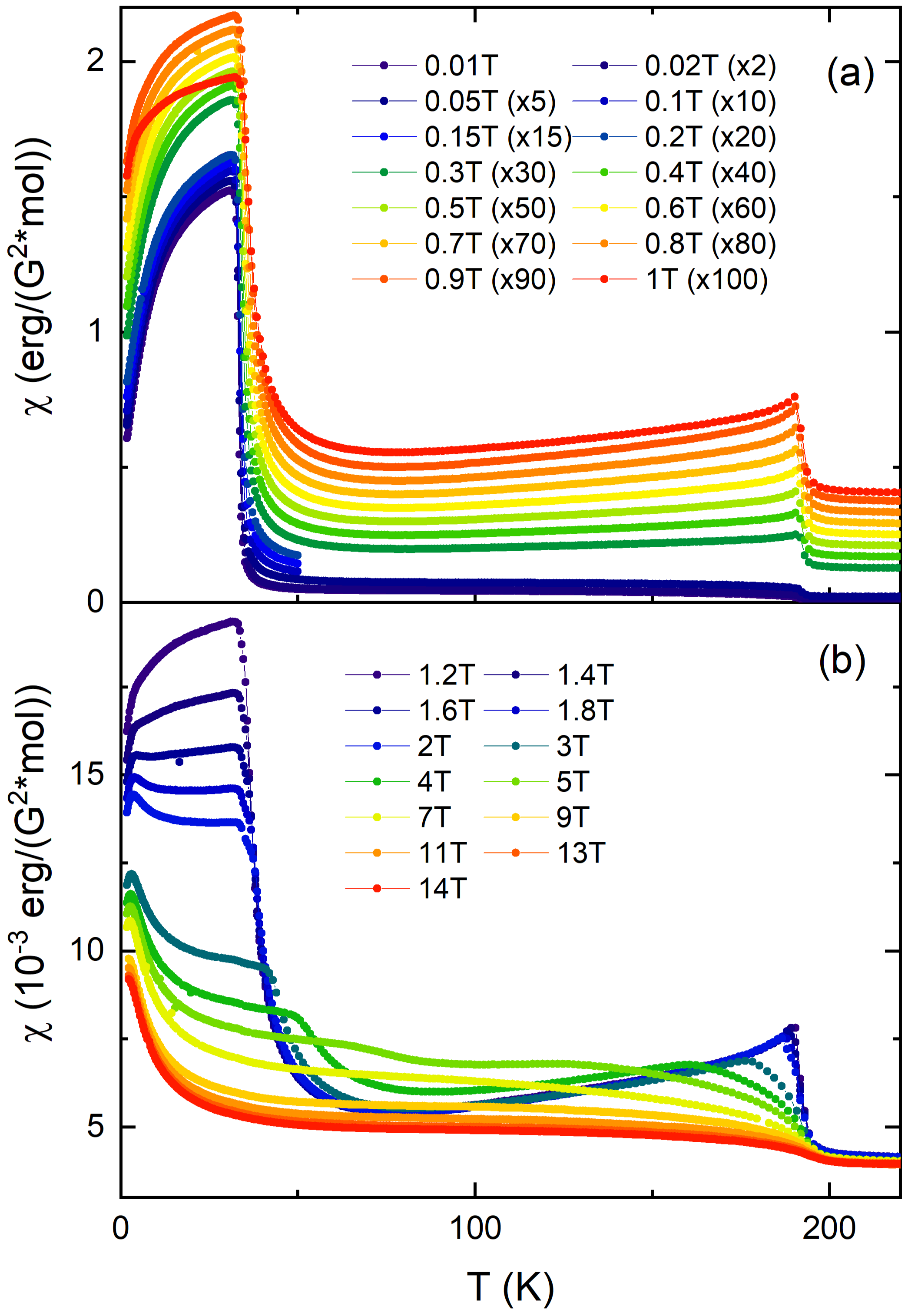}
\caption{Static magnetic susceptibility vs. temperature of \sco\ at different magnetic fields $B||a$ axis. Data in (a) have been multiplied as explained in the legend in order to enhance comprehension.}\label{SM_chia_all}
\end{figure}

\begin{figure}[hbt]
\includegraphics[width=\columnwidth,clip]{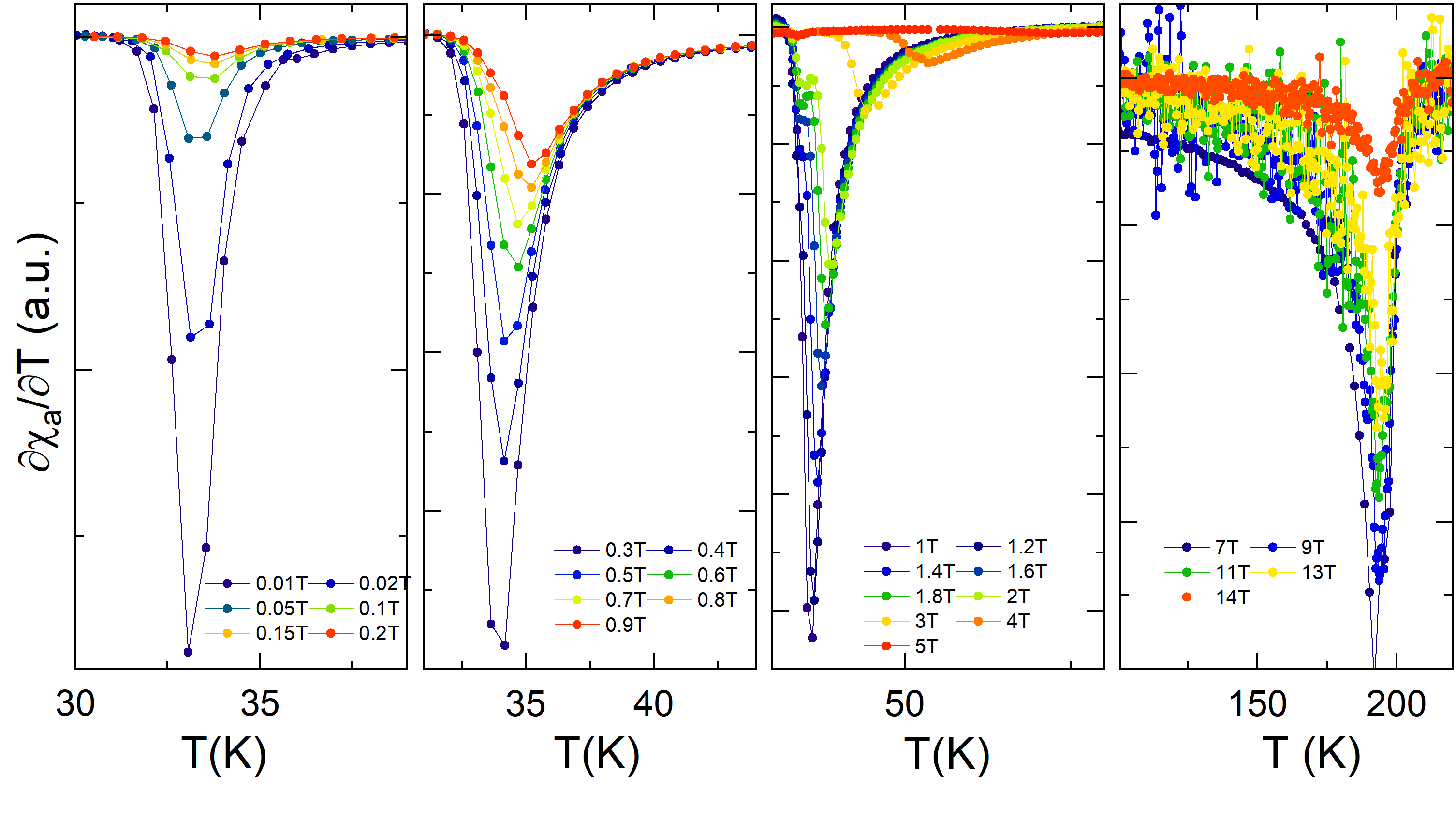}
\caption{Temperature derivative of the static magnetic susceptibility, $\partial \chi_a/\partial T$ for various fields $B||a$ axis. The anomaly positions have been used to construct the magnetic phase diagram.}\label{SCO_MBa}
\end{figure}

\begin{figure}[hbt]
\includegraphics[width=\columnwidth,clip]{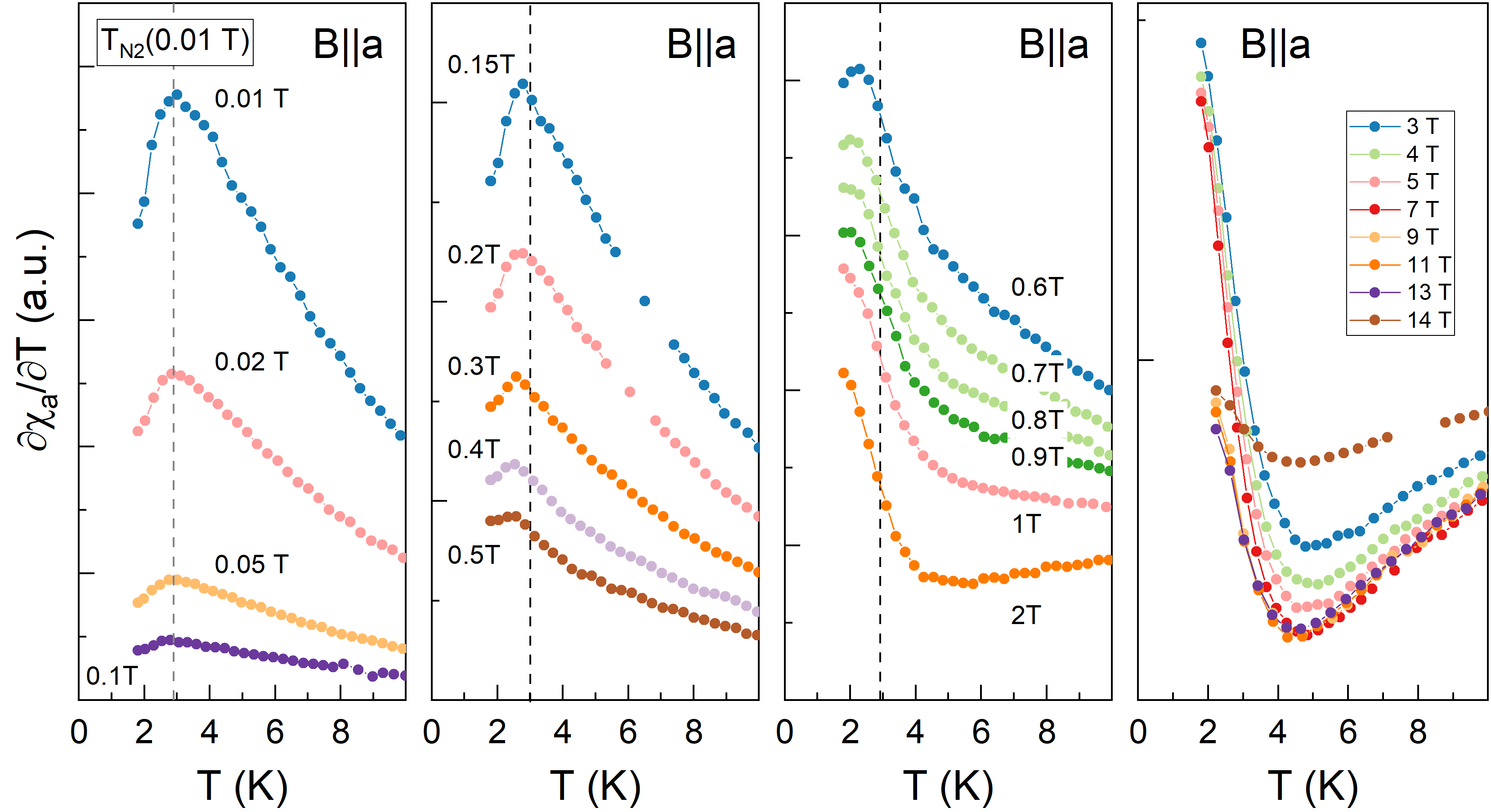}
\caption{Temperature derivative of the static magnetic susceptibility, $\partial \chi_a/\partial T$ for various fields $B||a$ axis, at low temperatures. The anomaly positions have been used to construct the magnetic phase diagram.}\label{SCO_MBa}
\end{figure} 

\begin{figure}[hbt]
\includegraphics[width=0.8\columnwidth,clip]{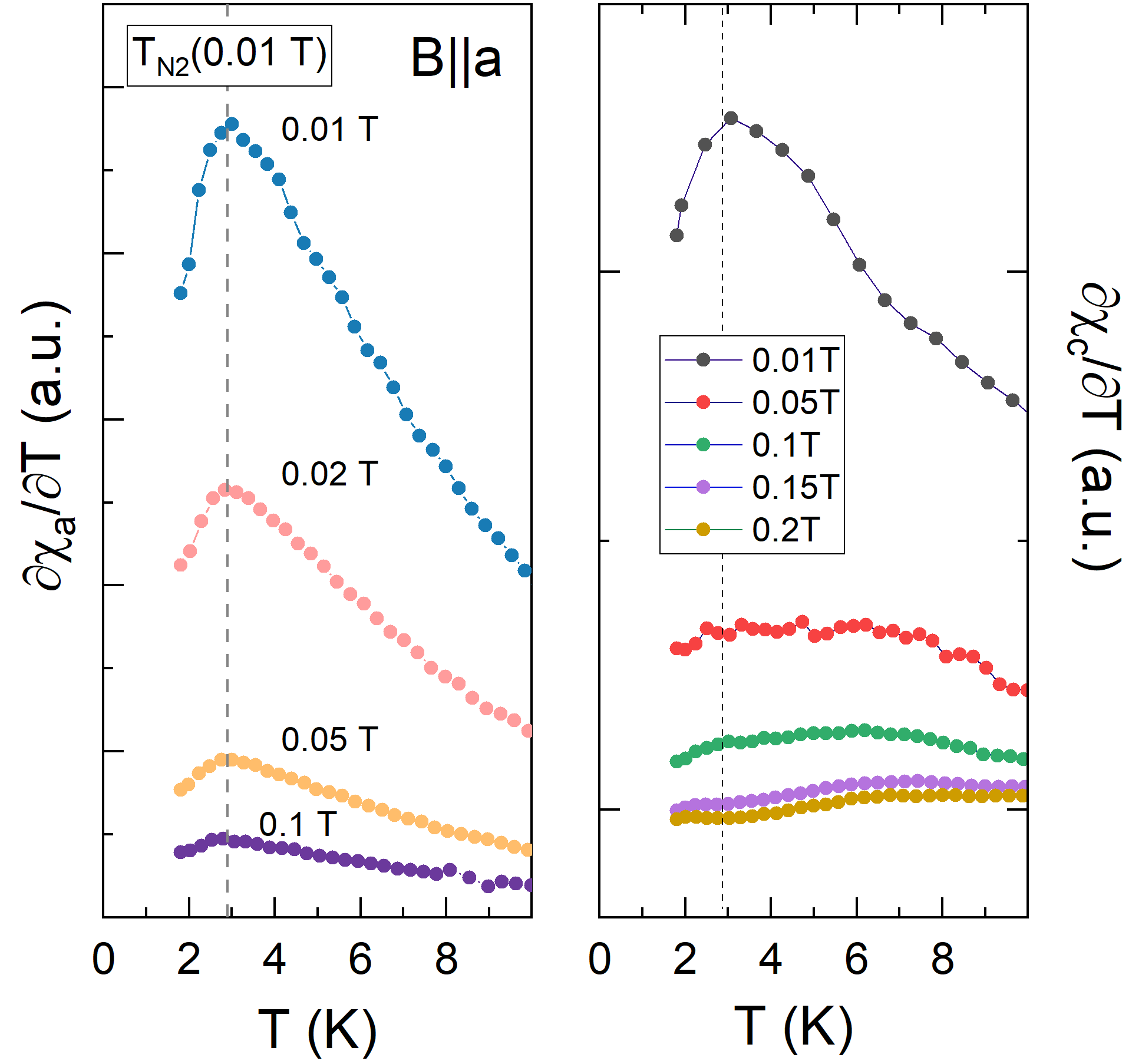}
\caption{Temperature derivative of the static magnetic susceptibilities, $\partial \chi_i/\partial T$ for various fields (a) $B||a$ axis, and (b) $B||c$ axis, at low temperatures. The dashed line marke the maximum for $B=10$~mT $||a$ in both plots. }\label{SM_Tn2b}
\end{figure} 

\begin{figure}[hbt]
\includegraphics[width=0.8\columnwidth,clip]{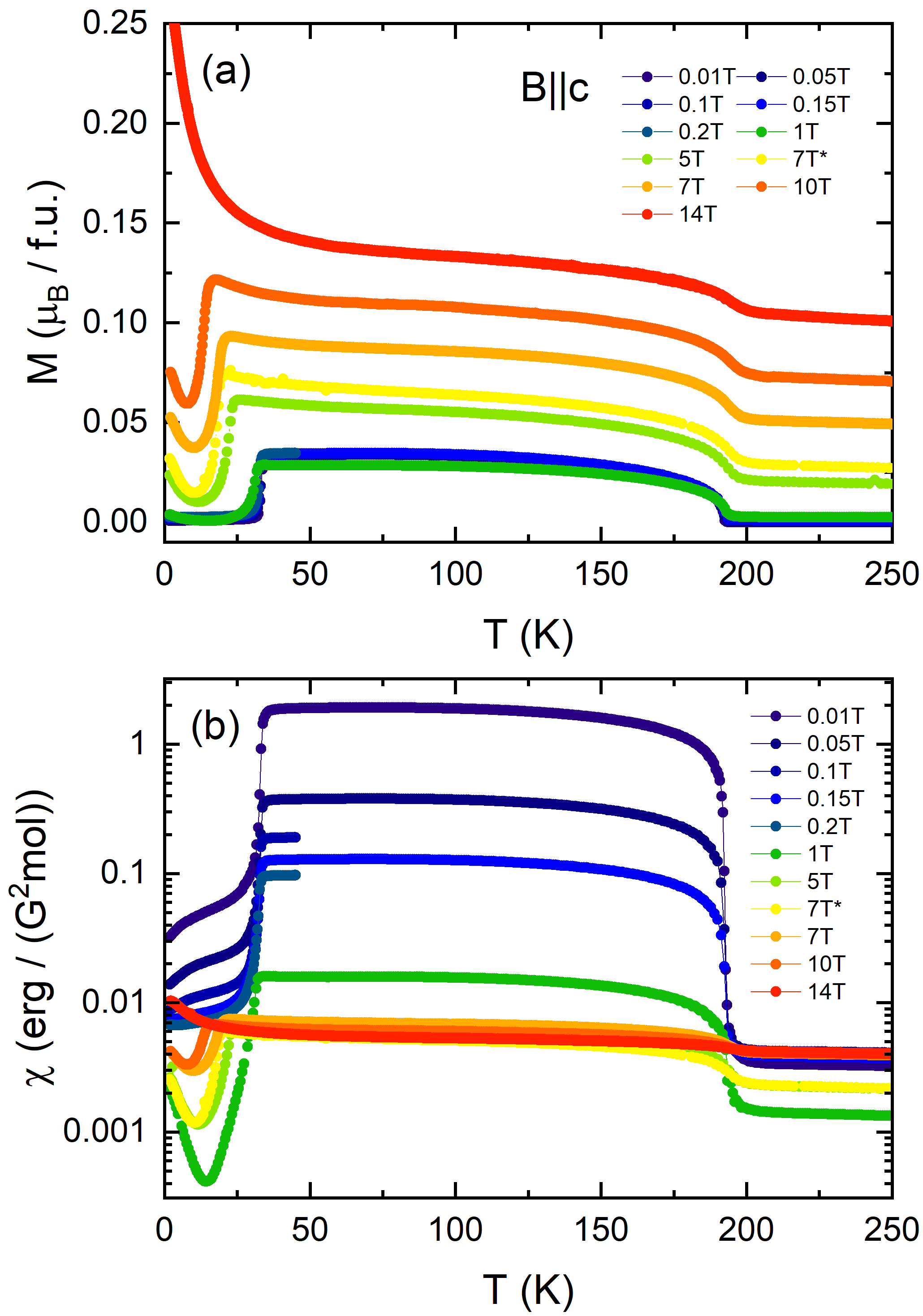}
\caption{Temperature dependence of (a) the magnetisation and (b) the static magnetic susceptibility of \sco\ at various magnetic fields $B||c$ axis. Data in (b) are shown on a log-scale for better visibility.}\label{SCO_MBc}
\end{figure} 

\begin{figure}[hbt]
\includegraphics[width=0.8\columnwidth,clip]{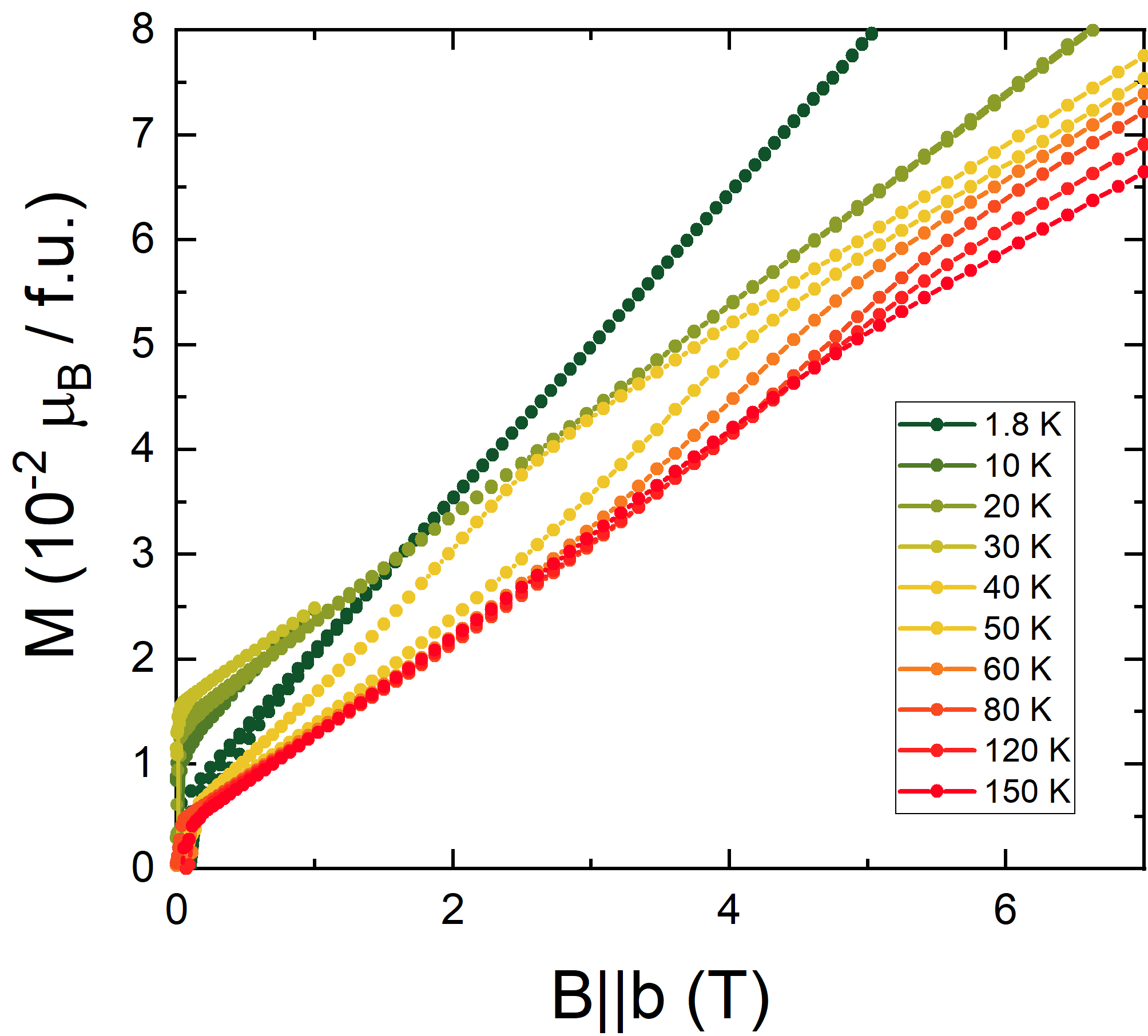}
\caption{Isothermal magnetization for $B||b$ axis at various temperatures. Derivatives are shown in Fig.~S4a.}\label{SM_MBb}
\end{figure}




\end{document}